%% file: number_counts_forecasts.tex
\documentclass[useAMS,usenatbib]{mn2e}
\pdfoutput=1
\usepackage{times}  
\usepackage{listings}
\usepackage{graphics}
\usepackage{subfigure}
\usepackage{graphicx}
\usepackage{amssymb}
\usepackage{amsmath}
\usepackage{aas_macros}
\usepackage{verbatim}
\usepackage{textgreek}
\usepackage{color}
\usepackage{url}
\usepackage{multirow}
\usepackage[dvipsnames]{xcolor}

\long\def\symbolfootnote[#1]#2{\begingroup%
\def\thefootnote{\fnsymbol{footnote}}\footnote[#1]{#2}\endgroup} 

\newcommand{\euclid}{\emph{Euclid}}

\newcommand{\galacticus}{\textsc{Galacticus}}
\newcommand{\galform}{\textsc{Galform}}

\newcommand{\cloudy}{\textsc{Cloudy}}

\newcommand{\halpha}{{\rm H\alpha}}
\newcommand{\Halpha}{{\rm H\alpha}}

\newcommand{\Mpc}{{\,\rm Mpc}}

\newcommand{\ergPerSecondPerCM}{{\rm erg}\,{\rm s}^{-1}{\rm cm}^{-2}}
\newcommand{\ergPerSecond}{{\rm erg}\,{\rm s}^{-1}}
\newcommand{\Msol}{{\rm M_{\odot}}}
\newcommand{\angstrom}{{\rm \AA}}

\newcommand{\emline}[1]{{\rm [#1]}}

\input{epsf} \title{Predicting $\boldsymbol\Halpha$ emission line galaxy counts for future galaxy\\ redshift surveys}

\author[Merson {\it et al.}]
       {\parbox[h]{\textwidth}{Alexander~Merson$^{1,2}$\thanks{E-mail:
       alex.i.merson@jpl.nasa.gov}, Yun~Wang$^{2,3}$, Andrew~Benson$^4$, Andreas Faisst$^2$,\\ Daniel Masters$^{1,2}$, Alina~Kiessling$^1$ \& Jason~Rhodes$^1$ }     
  \vspace*{10pt}\\
  \noindent$^1$Jet Propulsion Laboratory, California Institute of Technology, 4800 Oak Grove Drive, Pasadena, CA 91109, USA\\
  $^2$IPAC, Mail Code 314-6, California Institute of Technology, 1200 East California Boulevard, Pasadena, CA 91125, USA\\
$^3$Homer L. Dodge Department of Physics \& Astronomy, University of Oklahoma, 440 W Brooks Street, Norman, OK 73019, USA\\
$^4$Carnegie Observatories, 813 Santa Barbara Street, Pasadena, CA 91101, USA}
\date{}

\pagerange{\pageref{firstpage}--\pageref{lastpage}}
\pubyear{\the\year}

\begin{document}

\defcitealias{Ferrara99}{F99}
\defcitealias{Calzetti00}{C00}
\defcitealias{Charlot00}{CF00}
\defcitealias{Colbert13}{C13}
\defcitealias{Mehta15}{M15}

\maketitle
\title{Predictions for emission-line surveys}
\label{firstpage}

\begin{abstract}

Knowledge of the number density of $\Halpha$ emitting galaxies is vital for assessing the scientific impact of the Euclid and WFIRST missions. In this work we present predictions from a galaxy formation model, \galacticus, for the cumulative number counts of $\Halpha$-emitting galaxies. We couple \galacticus{} to three different dust attenuation methods and examine the counts using each method. A $\chi^2$ minimisation approach is used to compare the model predictions to observed galaxy counts and calibrate the dust parameters. We find that weak dust attenuation is required for the \galacticus{} counts to be broadly consistent with the observations, though the optimum dust parameters return large values for $\chi^2$, suggesting that further calibration of \galacticus{} is necessary. The model predictions are also consistent with observed estimates for the optical depth and the $\Halpha$ luminosity function. Finally we present forecasts for the redshift distributions and number counts for two Euclid-like and one WFIRST-like survey. For a Euclid-like survey with redshift range $0.9\leqslant z\leqslant 1.8$ and $\Halpha+\emline{NII}$ blended flux limit of $2\times 10^{-16}\ergPerSecondPerCM$ we predict a number density between 3900--4800 galaxies per square degree. For a WFIRST-like survey with redshift range $1\leqslant z\leqslant 2$ and blended flux limit of $1\times 10^{-16}\ergPerSecondPerCM$ we predict a number density between 10400--15200 galaxies per square degree.
\end{abstract}



\section{Introduction}
\label{sec:intro}

Explaining the observed accelerated expansion of the Universe
remains one of the most prominent questions of modern cosmology. 
The simplest explanation is that the expansion is being driven
by a negative-pressure component, referred to as \emph{dark
  energy}, thought to dominate the energy budget of the Universe. 
Whilst the majority of previous cosmological measurements
support the existence of dark energy, the uncertainties on these
measurements are too large to conclusively rule out alternative
theories (such as a modification of general relativity).
Distinguishing between dark energy and alternative theories is
possible given high precision measurements of the expansion history 
of the Universe and the growth rate of cosmic large-scale structure
\citep[e.g.][]{Albrecht06, Guzzo08, Wang08a, Wang08b}. 

Recent and upcoming cosmological galaxy surveys have therefore been designed 
to obtain these measurements using a set of complementary cosmological 
probes. Amongst these probes are \emph{baryon acoustic oscillations} 
(BAO), which can be used as a standard ruler to probe the expansion 
history of the Universe \citep{Blake03,Seo03}, and \emph{redshift-space 
distortions} (RSDs), which are sensitive to the growth rate of the 
large-scale structure \citep{Kaiser87,Song09}. Although both of these 
probes have been successfully measured with existing galaxy surveys 
\citep[e.g.][]{Peacock01, Cole05, Eisenstein05, Guzzo08}, the 
precision demanded to adequately test the nature of dark energy requires 
substantially larger galaxy samples, thus driving the need for
deep, wide-field surveys. 

Two such cosmological surveys, the ESA-led {\euclid} mission \citep{Laureijs11,Vavrek16} and NASA's \emph{Wide Field Infrared Survey Telescope} \citep[WFIRST,][]{Dressler12, Green12, Spergel15}, aim to measure the expansion history and growth rate of cosmic structure by using near-IR grism spectroscopy to target tens of millions of emission line galaxies (ELGs). These missions will predominantly target $\Halpha$ emitting galaxies, over the approximate redshift range $0.9\lesssim z \lesssim 2$. WFIRST will observe $\halpha$ at $1\lesssim z\lesssim 2$. Euclid is expected to probe a similar redshift range of $0.9\lesssim z\lesssim 1.8$. Note, however, that due to additional detection of [OIII] emission lines, both the Euclid and WFIRST missions will in fact be able to probe to $z\gtrsim 2$, though we will limit our focus here to the $\Halpha$ emission line. The Euclid wide-area survey will cover 15,000 square degrees to an $\halpha$ line flux limit of $2-3\times 10^{-16}\,\ergPerSecondPerCM$, while the WFIRST survey will cover $\sim2200$ square degrees to a fainter $\halpha$ line flux limit of $1\times 10^{-16}\,\ergPerSecondPerCM$. The smaller area and greater depth of WFIRST will be highly complementary to the larger area and shallower depth of Euclid. Note that these missions will also incorporate multi-band photometric surveys, which will exploit weak gravitational lensing as an additional probe of dark energy. The photometric component is expected to probe a similar effective volume (for example, see \citealt{Orsi10} for a comparison of the effective volumes of $\Halpha$ and photometrically selected surveys).

Assessing the expected performance of these future surveys is essential for optimising their design. The ability of galaxy surveys to achieve the desired precision, often quantified using a \emph{figure-of-merit}, is dependent upon the observed number density of the type of galaxies being targeted. Therefore, knowledge of the number density of different galaxies, as a function of redshift, is crucial in both the planning and evaluation of a survey strategy. As such, much effort has been invested in using existing emission line galaxy surveys to estimate the number of $\Halpha$ emitters that Euclid and WFIRST are likely to observe \citep[for example,][]{Colbert13, Mehta15, Pozzetti16,Valentino17}. 

Perhaps the most appropriate existing survey to date for making such estimates is the Wide Field Camera 3 Infrared Spectroscopic Parallels survey \citep[WISP,][]{Atek10,Atek11}, which uses the G141 ($1.2-1.7\mu m$, $R\sim130$) and G102 ($0.8-1.2\mu m$, $R\sim210$) grisms on the \emph{Hubble Space Telescope} Wide Field Camera 3 (HST WFC3) to acquire slit-less grism spectroscopy and imaging for galaxies in several hundred high-latitude fields. The wavelength ranges of the blue and red grisms on the WFC3 imply that the WISP survey is able to directly detect the $\Halpha$ emission line for galaxies out to redshifts of $z\lesssim1.5$. Although the total area of WISP is relatively small (on the order of 0.3 square degrees at the end of HST Cycle 23), the implementation of the WISP survey shares numerous commonalities with the proposed implementations for Euclid and WFIRST. As such, WISP is a very useful test-bed for predicting the number counts of emission line galaxies, the accuracy of redshift measurement, as well as the selection function and completeness of proposed survey strategies \citep{Mehta15}.

Measurements of the number density of $\Halpha$ emitters seen in the WISP survey have recently been made by \citet{Colbert13} and \citet{Mehta15}, who examined the cumulative $\Halpha$ galaxy number counts over the redshift range $0.7<z<1.5$. Note that neither \citet{Colbert13} nor \citet{Mehta15} made any flux corrections for dust attenuation or contamination due to the \emline{NII} ($6548{\rm \AA}+6584{\rm \AA}$) doublet. Using 29 separate WISP fields covering a combined area of 0.037 square degrees, \citet{Colbert13} estimated a number density of 6,700 ${\rm deg}^{-2}$ for $\Halpha$-emitting galaxies above a $\Halpha+[{\rm NII}]$ blended flux limit of $2\times10^{-16}\ergPerSecondPerCM{}$. Corrections due to incompleteness were modelled by simulating the line extraction process end-to-end. We refer the reader to \citet{Colbert13} for further details. Adopting a larger area of 52 WISP fields, covering a combined area of approximately 0.051 squared degrees, \citet{Mehta15} estimated a number density of 6,000 ${\rm deg}^{-2}$ for $\Halpha$-emitting galaxies above the same blended flux limit. When the blended flux limit was reduced to $1\times10^{-16}\ergPerSecondPerCM{}$, both \citet{Colbert13} and \citet{Mehta15} estimated a number density of 15,000 ${\rm deg}^{-2}$. Updated galaxy counts, using a larger number of WISP fields are expected in the near future (Bagley \textit{et al.}, in prep.). 

In their analysis, \citet{Mehta15} also estimated the WISP $\Halpha$ counts out to $z<2$ by firstly measuring the bivariate $\Halpha$--[OIII] line luminosity function for $0.7<z<1.5$ and then using the observed correlation between the $\halpha$ and [OIII] luminosities to predict the $\Halpha$ counts from the [OIII]-only line luminosity function (see \citealt{Mehta15} for further details). Over the redshift range $0.7<z<2$ \citet{Mehta15} find the number density of $\Halpha$-emitters to be approximately 11,000 ${\rm deg}^{-2}$ for a $\Halpha+[{\rm NII}]$ blended flux limit of $2\times10^{-16}\ergPerSecondPerCM{}$ and approximately 32,000 ${\rm deg}^{-2}$ for a limit of $1\times10^{-16}\ergPerSecondPerCM{}$ (again they made no correction for dust attenuation).

Attempts to describe the $\Halpha$ line luminosity functions (and hence the number counts) using empirically motivated models have recently been presented in \citet{Pozzetti16}, who compared three different models that were fitted to the $\Halpha$ luminosity functions measured from WISP, the ground-based narrow-band High-z Emission Line Survey \citep[HiZELS,][]{Geach08,Sobral09}, as well as datasets obtained using the Near Infrared Camera and Multi-Object Spectrometer \citep[NICMOS,][]{Yan99,Shim09}. In this work we present the cumulative $\Halpha$ count predictions from an open-source semi-analytical galaxy formation model. Our work is complementary to that of \citet{Pozzetti16}, though has more predictive power since we use simulations that incorporate a physical model of galaxy formation and are calibrated using the observational data. We compare the predicted counts to the WISP estimates, in particular the estimates from \citet{Mehta15}. In Section~\ref{sec:galaxy_formation_model} we introduce the model and describe the calculation of emission line properties. After this we present the model predictions, starting in Section~\ref{sec:number_counts} with a comparison with the WISP number counts. In this section we perform a $\chi^2$ minimisation to calibrate the parameters for several dust attenuation methods, and briefly examine the impact of fixing the $\emline{NII}/\Halpha$ line ratio as well as the impact of cosmic variance. In Section~\ref{sec:luminosity_functions} we compare the model $\Halpha$ luminosity function to observed luminosity functions at several redshift epochs from WISP and HiZELS. In Section~\ref{sec:forecasts} we present forecasts for the number of $\Halpha$-emitting galaxies expected from two Euclid-like surveys and one WFIRST-like survey. Finally we summarise our findings in Section~\ref{sec:conclusions}. The cosmology assumed throughout this work, a baryon matter density $\Omega_{{\rm b}} = 0.045$, a total matter density $\Omega_{{\rm m}} = \Omega_{{\rm b}} + \Omega_{{\rm CDM}} = 0.25$, a dark energy density $\Omega_{{\rm \Lambda}} = 0.75$ and a Hubble constant $H_0 = 100h \,{\rm km\,s}^{-1}\Mpc^{-1}$ where $h = 0.73$, is chosen to match the cosmology used in the N-body simulation that was used (see Section~\ref{sec:nbody_simulation}).

\section{Galaxy formation model}
\label{sec:galaxy_formation_model}

In this work we adopt the open source {\galacticus}\footnote{Here we use version 0.9.4 of {\galacticus}, which is publicly available from: \url{https://sites.google.com/site/galacticusmodel}. The Mercurial hash I.D. for the particular revision used is: 4787d94cd86e.} semi-analytical galaxy formation model \citep{Benson12}.

\subsection{Overview of the \galacticus{} model}
\label{sec:galacticus}

Like other such galaxy formation models, \galacticus{} is designed to construct and evolve a population of galaxies within a merging hierarchical distribution of dark matter halos. \galacticus{} does this by solving a set of coupled ordinary differential equations (ODEs) that govern the various astrophysical processes affecting the baryonic matter within the halos. This includes, for example, the rate of radiative gas cooling, the quiescent star formation rate, the chemical enrichment of the stellar and gaseous components, as well as the regulation of feedback processes from supernovae and active galactic nuclei. Various galaxy properties, such as the stellar mass or cold gas metallicity, at any given epoch can the be computed by calling the ODE solver within \galacticus{} that integrates the set of equations forward to the desired time with a user-specified level of precision. Galaxy mergers are treated in \galacticus{} as impulsive events that occur instantaneously and interrupt the evolution of the galaxy properties by the ODE solver. (Interruption of the evolution is allowed by the ODE solver.) Once the merger event is completed the updated galaxy properties are passed to the ODE solver to continue their evolution. Convolving the star formation history of a galaxy with a specified single stellar population synthesis (SPS) model, with the assumption of a stellar initial mass function (IMF), allows one to construct an SED for each galaxy, which can be sampled under different filter transmission curves to obtain photometric luminosities. By default \galacticus{} uses the \textsc{Flexible Stellar Population Synthesis} code of \citet{Conroy10} and assumes a \citet{Chabrier03} IMF. 
 
\galacticus{} is calibrated to reproduce various observational statistics of the galaxy population, with particular emphasis on observations of the population in the local Universe, which are the most tightly constrained. A principal statistic used for calibration is the present day galaxy stellar mass function as measured from the Sloan Digital Sky Survey (SDSS) by \citet{Li09}. Further details of the calibration of the model, as well as comparisons between \galacticus{} and several other semi-analytical models, can be found in \citet{Knebe15}. In this work we use the default \galacticus{} parameter set and examine only the impact of varying the strength of dust attenuation.

\subsection{N-body Simulation}
\label{sec:nbody_simulation}

\galacticus{} is able to model the collapse and merging of dark matter halos using the Press-Schechter formalism or work with halo merger trees extracted from a cosmological N-body simulation. Here we adopt the latter approach and work with halos extracted from the \emph{Millennium Simulation} \citep{Springel05}. We adopt the Millennium Simulation as it is currently one of the largest volume simulations for which \galacticus{} has been calibrated.

Based upon the ${\rm \Lambda}$CDM cosmology, the Millennium Simulation uses $2160^3$ particles to follow the hierarchical growth of dark matter structures in a cubical volume with side length of $500h^{-1}\Mpc$. The simulation follows the growth of structures from high redshift ($z=127$) through to the present day with particle and velocity information stored for 63 snapshots at fixed epochs spaced approximately logarithmically in expansion factor between $z=20$ and $z=0$ \citep{Lemson06}. For each of these snapshots the \textsc{Subfind} halo-finding algorithm \citep{Springel01} was used to identify bound halo and sub-halo structure down to a resolution limit of 20 particles. Given the particle mass of $8.6\times10^{8}h^{-1}\Msol$ this yields a minimum halo mass of approximately $1.72\times10^{10}h^{-1}\Msol$. The cosmology adopted in the simulation is: a baryon matter density $\Omega_{{\rm b}} = 0.045$, a total matter density $\Omega_{{\rm m}} = \Omega_{{\rm b}} + \Omega_{{\rm CDM}} = 0.25$, a dark energy density $\Omega_{{\rm \Lambda}} = 0.75$, a Hubble constant $H_0 = 100h \,{\rm km\,s}^{-1}\Mpc^{-1}$ where $h = 0.73$, a primordial scalar spectral index $n_{\rm s}=1$ and a fluctuation amplitude $\sigma_{8}=0.9$. This cosmological parameter set is consistent with the first year results from the Wilkinson Microwave Anisotropy Probe \citep{Spergel03}. Note that we do not expect our examination of the number density of $\Halpha$ emitters to be particularly sensitive to our choice of cosmology. The scatter in the observed number counts of $\Halpha$ emitters is much larger than any differences we would expect due to cosmology. Furthermore, any uncertainties in the predicted counts due to cosmology will be negligible compared to our ignorance of the underlying galaxy formation physics.

\subsection{Lightcone construction}
In order to  make predictions for the number counts of $\Halpha$-emitting galaxies we need to build a \emph{lightcone} catalogue, whereby only structures that intersect the past lightcone of an observer are included in the catalogue. The \galacticus{} model uses the lightcone construction methodology of \citet{Kitzbichler07} to identify  dark matter halos that intersect the lightcone of the observer. We build a lightcone of 4 square degrees (an area comparable to the COSMOS field) that spans the redshift range $z\simeq0.5$ to $z\simeq3$. This redshift range was chosen to encompass the approximate redshift ranges that will be probed by the emission line selected spectroscopic components of the Euclid and WFIRST missions (though in this work we will only be considering redshifts for which the $\Halpha$ line is observable). Once the halos are identified, \galacticus{} will process only the merger trees that contain these halos. As such, \galacticus{} will simply output all of the galaxies that occupy the halos in this volume and does not apply any radial selection criteria, which can be applied in post-processing.

\begin{figure}
  \centering
  \includegraphics[width=0.48\textwidth]{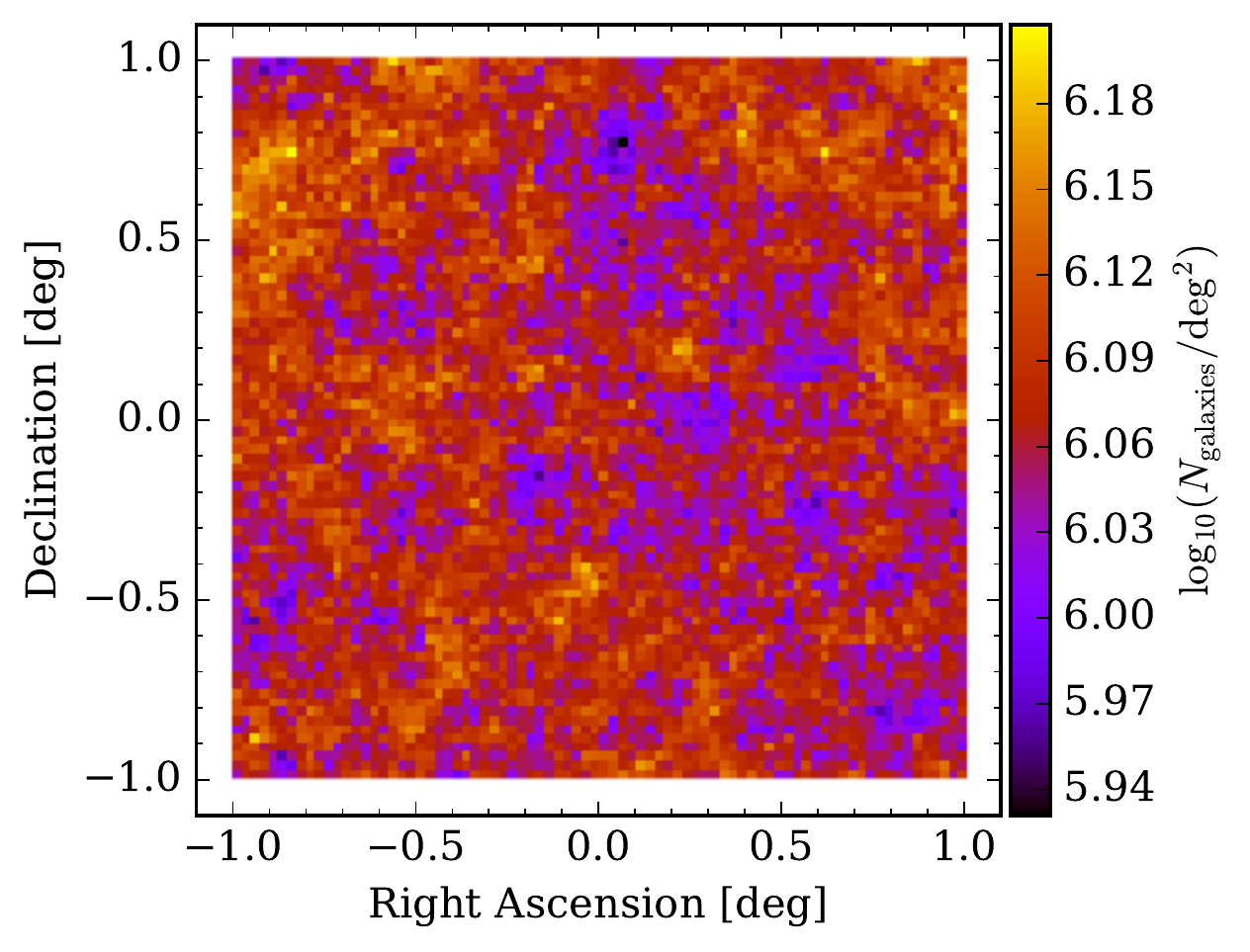}
  \caption{Projected galaxy counts in \galacticus{} $2\times2$ square degrees lightcone. The pixel colours indicate the total number of \galacticus{} galaxies, per square degree, between $0.5\lesssim z\lesssim3$ in that small patch on the sky. Note that no radial selection has been applied to the galaxies. The pixels clearly show the presence of projected large-scale structure within the lightcone volume.}
  \label{fig:sky_distribution}
\end{figure}

We show in Fig.~\ref{fig:sky_distribution} the projected distribution of \galacticus{} galaxies on the sky for our 4 square degrees lightcone. The presence of projected large-scale structure in the lightcone is clear in the number of galaxies per pixel. 

\subsection{Emission line modelling}
\label{sec:galacticus_emlines}

Emission line luminosities are computed for \galacticus{} galaxies by following a methodology similar to that of \citet{Panuzzo03}, who used the photo-ionisation code {\cloudy} \citep{Ferland13}\footnote{\citet{Ferland13} presents version 13.03 of {\cloudy}. The version used here is version 13.04.} to compute the luminosity of various emission lines as a function of the number of ionising photons for various species (HI, HeI and OII), the metallicity of the inter-stellar medium (ISM), the hydrogen gas density and the volume filling factor of HII regions. Our methodology is as follows:

\begin{enumerate}
\item We first generate a grid of \cloudy{} models of HII regions spanning a range of hydrogen gas densities, gas metallicities and ionising luminosities for the HI, HeI and OII ionising continua. For each of the given values for the ionising luminosities in the three continuua we provide \cloudy{} with an input ionisation spectrum consisting of a combination of three black body spectra. The first black body spectrum is designed to span the HeI ionising wavelength region, whilst the second is designed to span the HI ionising region. The third black body spectrum spans the non-ionising region, i.e. everything at wavelengths longer than the HI Lyman ionisation edge. The temperatures and normalisations of these three spectra are derived as detailed in Appendix A of \citet{Panuzzo03}. From each \cloudy{} model we extract the luminosity for each emission line that is of interest. These luminosities are then stored in a library so that this first step need only be performed once, or whenever additional emission lines are required.

\item For each \galacticus{} galaxy an integrated stellar spectrum is computed by integrating simple stellar population models based upon the star formation and metallicity histories of the galaxy, as well as the assumed stellar initial mass function. The ionising luminosities in HI, HeI and OII are computed by then integrating over the stellar spectrum. This is similar to the approach taken by \citet{Orsi08}.

\item  The characteristic hydrogen gas density, $\rho_{\mathrm{H}}$, of the ISM in the galaxy disc or spheroid (bulge) is computed as,
\begin{equation}
\rho_{\mathrm{H}} \propto \frac{M_{\mathrm{gas}}f_{\mathrm{H}}}{\mu m_{\rm H}} \frac{1}{4 \pi r^3},
\label{eq:densityHydrogen}
\end{equation}
where $M_{\mathrm{gas}}$ is the mass of the ISM in the galaxy disc or spheroid, $f_{\mathrm{H}}$ is the mass fraction of hydrogen (assumed to take the solar value of 0.707), $\mu$ is the atomic mass, $m_{\rm H}$ is the mass of hydrogen in atomic mass units, and $r$ is the radius of the disc or spheroid. In \galacticus{} emission line luminosities are computed for the galaxy disc and spheroid separately. These two components are summed to give the luminosity of the entire galaxy.

\item  The number of HII regions, $N_{\rm HII}$, in the galaxy disc or spheroid is estimated according to,
\begin{equation}
N_{\rm HII} \sim \psi\frac{\tau_{\rm HII}}{M_{\rm HII}},
\end{equation}
where $\psi$ is the instantaneous star formation rate in the disc or spheroid, $\tau_{\rm HII}$ is the typical lifetime of an HII region and $M_{\rm HII}$ is the typical mass of an HII region. We assume $\tau_{\rm HII}\sim 1{\rm Myr}$ and $M_{\rm HII}\sim 7.5\times10^3{\rm M_{\odot}}$, consistent with the values adopted by \citet{Panuzzo03}. Under the assumption that all HII regions are identical, we divide the HI, HeI and OII ionising luminosities by $N_{\rm HII}$ to obtain the ionising luminosities for a single HII region.

\item Taking the hydrogen density, $\rho_{\mathrm{H}}$, the metallicity of the ISM (computed by \galacticus{}) and the ionising luminosities per HII region, we interpolate over the grid of \cloudy{} models to obtain the emission line luminosity per HII region. We then multiply this by the number of HII regions in the galaxy disc or spheroid, $N_{\rm HII}$, to obtain the total emission line luminosity for that component. 
\end{enumerate}

With this approach we are able to self-consistently compute luminosities for any emission line that is output by \cloudy{}. As such, \galacticus{} is a viable model for performing a wide range of studies of the properties of different emission line galaxies. In this first application we will focus only on the $\Halpha$ emission line. In planned  future work we will examine the predictions for several other lines, including the \emline{NII} and \emline{OIII}  lines, which will also be detected by Euclid and WFIRST\footnote{With a grism resolution of $R=461\lambda$, the WFIRST mission is expected to resolve the \emline{NII} doublet from the $\Halpha$ line for most galaxies.}.

\subsection{Attenuation by dust}
\label{sec:galacticus_dust}

In \galacticus{} attenuation of stellar and emission line luminosities due to dust is applied in post-processing. As such, \galacticus{} can be combined with any number of dust attenuation methods, ranging from simple dust screening methods \citep[e.g.][]{Calzetti00} to sophisticated radiative transfer codes, such as \textsc{GRASIL} \citep{Silva98}. In this work we consider three different dust attenuation methods: by \citet{Ferrara99}, \citet{Charlot00} and a simple \citet{Calzetti00} law. Calibration of the parameters used in these dust methods, using $\chi^2$ minimisation to match the WISP number counts, is presented in Section~\ref{sec:dust_calibration}.

\subsubsection{The \citet{Charlot00} method}
\label{sec:CharlotFall_method}

The simple model devised by \citet{Charlot00} is able to predict the attenuation due to scattering and absorption of light by interstellar dust using a basic description of the main features of the ISM. \citeauthor{Charlot00} assume that young stars are born in the centres of dense molecular clouds, which are ionised by stellar radiation to form ionized HII regions, surrounded by HI envelopes.

Based upon the \citet{Charlot00} method, we compute the dust-attenuated luminosity, $L^{\rm att}$, for a galaxy according to,
\begin{eqnarray}
L^{\rm att} &=& \left(L^{\rm int} - L^{\rm int}_{\rm recent}\right)\exp\left(-\tau^{\rm ISM}_{\lambda}\right)\nonumber\\
&+& L^{\rm int}_{\rm recent}\exp\left(-\tau^{\rm ISM}_{\lambda}\right)\exp\left(-\tau^{\rm MC}_{\lambda}\right),
\label{eq:CharlotFallLuminosity}
\end{eqnarray}
where $L^{\rm int}$ is the intrinsic (dust-free) luminosity for the galaxy, $L^{\rm int}_{\rm recent}$ is the intrinsic luminosity coming from recent star formation, $\tau^{\rm ISM}_{\lambda}$ is the optical depth of the diffuse ISM and $\tau^{\rm MC}_{\lambda}$ the optical depth of the molecular clouds. We assume that $L^{\rm int}_{\rm recent}$ corresponds to the luminosity coming only from stars formed within the past $10\,{\rm Myr}$. Since emission line luminosities correspond to recent star formation, we approximate $L^{\rm int}\approx L^{\rm int}_{\rm recent}$ and so for emission lines $L^{\rm att}\approx L^{\rm int}_{\rm recent}\exp\left(-\tau^{\rm ISM}_{\lambda}\right)\exp\left(-\tau^{\rm MC}_{\lambda}\right)$. 

The optical depth of the ISM, as a function of wavelength $\lambda$, is computed as,
\begin{equation}
\tau^{\rm ISM}_{\lambda}\propto\hat{{\tau}}^{\rm ISM}_{\rm V}\Sigma_{\rm gas}^{\rm metals}\left(\frac{\lambda}{\lambda_0}\right)^{-n},
\label{eq:opticalDepthISM}
\end{equation}
where $\hat{{\tau}}^{\rm ISM}_{\rm V}$ is an effective absorption optical depth of the ISM in the V-band ($5500{\rm \AA}$), $\Sigma_{\rm gas}^{\rm metals}$ is the central surface density of metals in the cold gas of the galaxy, $\lambda_0$ is a wavelength zero-point and $n$ is a power law exponent. The optical depth of the molecular clouds, as function of wavelength $\lambda$, is computed as,
\begin{equation}
\tau^{\rm MC}_{\lambda}=\hat{{\tau}}^{\rm MC}_{\rm V}\frac{Z_{\rm gas}}{Z_{\rm ISM}}\left(\frac{\lambda}{\lambda_0}\right)^{-n},
\label{eq:opticalDepthClouds}
\end{equation}
where $\hat{{\tau}}^{\rm MC}$ is an effective absorption optical depth for molecular clouds in the V-band, $Z_{\rm gas}$ is the cold gas metallicity of the galaxy and $Z_{\rm ISM}$ is the metallicity of the local ISM in the solar neighbourhood.

The parameters $\hat{{\tau}}^{\rm ISM}_{\rm V}$, $\hat{{\tau}}^{\rm MC}_{\rm V}$, $\lambda_0$ and $n$ are specified in the \citet{Charlot00} method. Following calibration against a set of local, UV-bright starburst galaxies, \citet{Charlot00} set the values of these parameters to $\hat{{\tau}}^{\rm ISM}_{\rm V}=0.5$, $\hat{{\tau}}^{\rm MC}_{\rm V}=1.0$, $\lambda_0=5500{\rm \AA}$ and $n=0.7$.

\subsubsection{The \citet{Ferrara99} library}
\label{sec:Ferrara_method}

The \citet{Ferrara99} method uses a set of Monte-Carlo simulations, designed to examine the scattering and absorption of light due to dust, to generate a library of dust attenuation curves as a function of various galaxy properties. Within a galaxy, the diffuse dust in the ISM is assumed to trace the distribution of stars. Given a particular extinction law, the \citet{Ferrara99} method will provide dust attenuation factors as a function of wavelength, galaxy
inclination, the ratio of spheroid to disc radial dust scale length, $r_e/h_R$, the ratio of dust to stellar vertical scale heights, $h_{z{\rm ,\,dust}}/h_{z{\rm ,\,stars}}$, and the V-band optical depth when looking face-on through the centre of
a galaxy, $\tau_{{\rm V}}^0$. All of these dust properties can be calculated directly from \galacticus{} predictions. For example, $\tau_{{\rm V}}^0$, is given by,
\begin{equation}
  \tau_{{\rm V}}^0 \propto \frac{M_{{\rm cold}}Z_{{\rm
        cold}}}{r^2_{{\rm disc}}},
  \label{eq:optical_depth}
\end{equation}
where $M_{{\rm cold}}$ is the mass of cold gas in the galaxy (both atomic and molecular gas), $Z_{{\rm cold}}$ is the metallicity of the cold gas and $r_{{\rm disc}}$ is the radius of the galactic disc. As such, the strength of attenuation due to dust at any particular wavelength for any particular \galacticus{} galaxy can be determined by interpolation over the \citet{Ferrara99} library. The advantage of this physically-motivated method therefore is that the dust attenuation varies self-consistently with other galaxy properties, such as the cold gas content and galaxy sizes.

Although the \citeauthor{Ferrara99} method is currently one of the standard attenuation options used in \galacticus{} post-processing, this approach does not include the attenuation due to much denser molecular clouds embedded within the ISM. Molecular clouds are regarded as the birth places of new stars and sites of ongoing star-formation, and so we would therefore expect them to be intense sources of emission lines. As such, it is necessary to either incorporate molecular clouds into the \citeauthor{Ferrara99} method (as has been done for instance in the \galform{} semi-analytical model, see for example \citealt{Cole00}, \citealt{Gonzalez-Perez13} and \citealt{Lacey16}). We can introduce nebular dust attenuation using the approach from the \citet{Charlot00} method described previously. We can compute the attenuated luminosity using Eq.~\ref{eq:CharlotFallLuminosity}, but substituting $\tau^{\rm ISM}_{\lambda}$ with the ISM optical depth computed in Eq.~\ref{eq:optical_depth}. Hereafter, when discussing the \citet{Ferrara99} method, we will refer to this modified method that includes attenuation from molecular clouds.

\subsubsection{The \citet{Calzetti00} law}
\label{sec:Calzetti_method}

Unlike the previous two methods, the \citet{Calzetti00} method employs a simpler empirical approach by modelling the attenuation assuming that all the dust forms an optically thick screen between the galaxy and the observer. This method, which was originally calibrated using observations of 39 near-by UV-bright starburst galaxies, simply provides a parametrised fit to the dust attenuation curve and requires little information with regards to other galaxy properties. For completeness we note that several recent studies of larger galaxy samples at both low and high redshift have found the law to be a poor description for dust attenuation at UV wavelengths \citep[e.g.][]{Noll09,Conroy10,Wild11,Buat11a,Buat11b,Buat12,Lee12,Reddy12}. However, despite this and despite its non-physical set-up, the ease of use of the \citet{Calzetti00} law has made it a popular choice in observational analyses. As such, parametrised fitting of the dust attenuation curve is a common treatment for dust attenuation and is used by a wide variety of similar screening methods, \citep[e.g.][]{Seaton79,Prevot84,Bouchet85,Fitzpatrick86,Cardelli89,Gordon03}, which typically differ only by their particular choice of parametrisation. We note that all such screen attenuation methods can be easily applied to \galacticus{} galaxies and that we have simply chosen the \citet{Calzetti00} method as a demonstration. 

\section{Comparison to observed number counts}
\label{sec:number_counts}

Before we make number counts predictions for Euclid and WFIRST, it is necessary to first compare the number counts predicted by \galacticus{} to existing observations and, where necessary, calibrate \galacticus{} to match these observations. In this section we will therefore compare the \galacticus{} counts to the observed counts from the WISP survey, as presented by \citet{Colbert13} and \citet{Mehta15}. In particular we will focus on comparing to the most recent results from \citet{Mehta15} over the redshift range $0.7<z<1.5$. We remind the reader that due to the upper wavelength limit of the HST WFC3 G141 grism, 1.2--1.7 microns, the $\Halpha$ line is only directly detectable with WFC3 out to $z\lesssim1.5$. When computing the galaxy counts from \galacticus{} we therefore only consider galaxies within the redshift range, $0.7<z<1.5$.

\subsection{Modelling \emline{NII} contamination}
\label{sec:modellingNII}

\begin{figure*}
  \centering
  \includegraphics[width=0.98\textwidth]{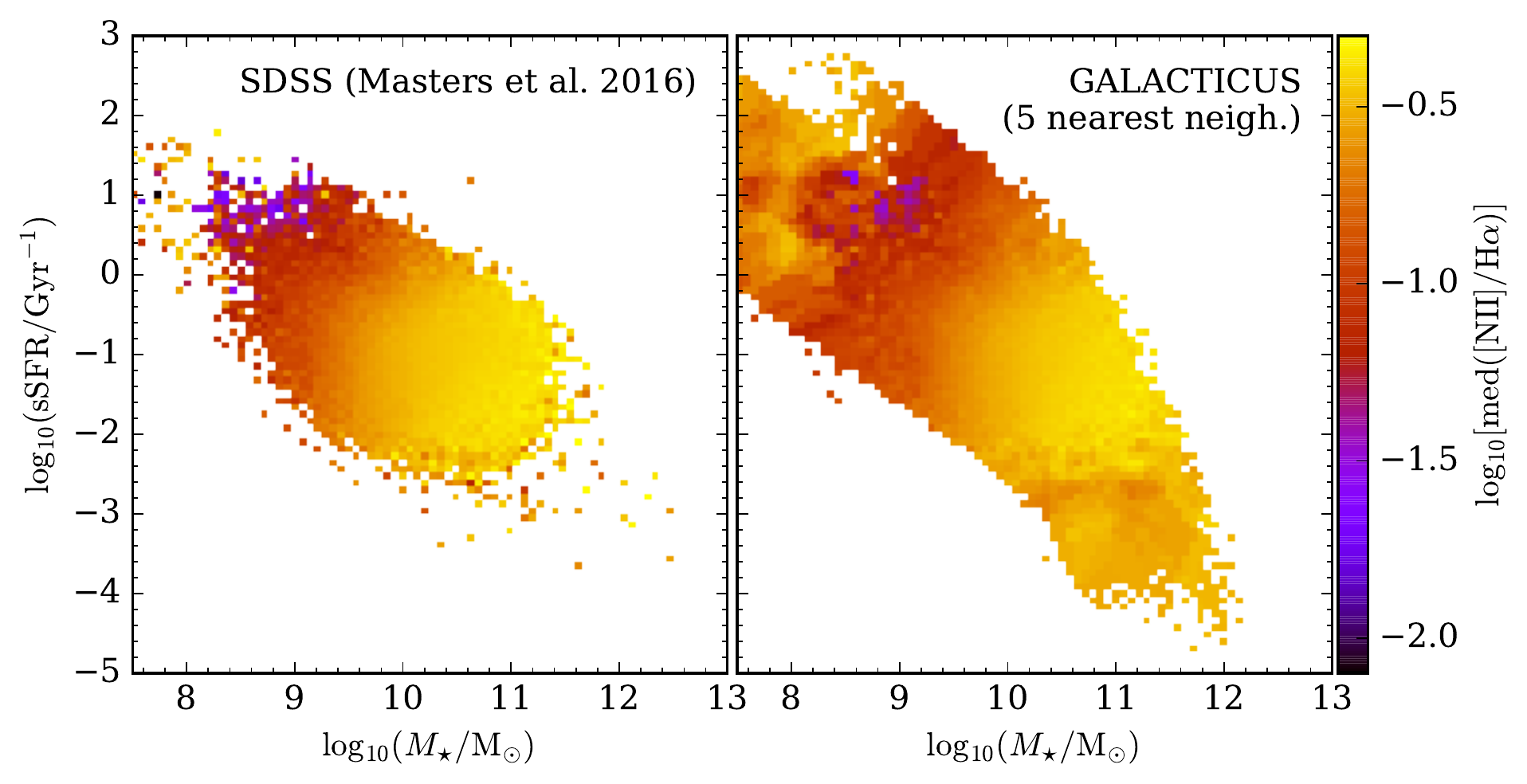}
  \caption{Distribution of the ${\rm \emline{NII}}/\Halpha$ line ratio in the stellar mass ($M_{\star}$) vs. specific star formation rate (sSFR) plane. The coloured pixels show the logarithm base-10 of the median  line ratio for the galaxies in that $M_{\star}$ vs. ${\rm sSFR}$ bin. In the left-hand panel we show the distribution for the SDSS galaxies in the catalogue presented in \protect\citet{Masters16}. In the right-hand panel we show the distribution obtained for the \galacticus{} lightcone when we assign the \galacticus{} galaxies with line ratios taken from the SDSS catalogue.  The \galacticus{} galaxies have redshifts within the range $0.7<z<1.5$ and we show those galaxies with an intrinsic (rest-frame), dust-free $\Halpha$ luminosity above $10^{40}\ergPerSecond$. For each \galacticus{} galaxy we first identify in $M_{\star}$ vs. ${\rm sSFR}$ space the 5 nearest neighbours from the SDSS catalogue and then assign the \galacticus{} galaxy a linear ratio equal to the median of the line ratios from these 5 SDSS galaxies.}
  \label{fig:sdss_vs_galacticus_line_ratios}
\end{figure*}

The galaxy counts presented by \citet{Mehta15} have not been corrected for dust attenuation or for contamination from \emline{NII}. Due to the resolution of the WFC3 grisms, WISP is unable to resolve the \emline{NII} doublet and de-blend it from the $\Halpha$ line. In their analysis, \citet{Colbert13} and \citet{Mehta15} assumed a fixed \emline{NII} contamination of 29 per cent, i.e. the $\Halpha$ flux, $f_{\rm \Halpha}$, is assumed to be $f_{\rm \Halpha}=0.71f_{\Halpha+\emline{NII}}$, where $f_{\Halpha+\emline{NII}}$ is the observed $\Halpha+\emline{NII}$ flux.

In \galacticus{} the luminosities for the \emline{NII} doublet lines can be computed by \cloudy{}. However, we find that the $\emline{NII}/\Halpha$ ratio computed using the luminosities obtained from \cloudy{} are typically an order of magnitude too small compared to $\emline{NII}/\Halpha$ ratios measured by \citet{Masters14} for a subset of WISP galaxies. Whilst the WISP galaxies presented by \citeauthor{Masters14} typically have $\emline{NII}/\Halpha$ between approximately 0.1 and 0.2, we find that the \cloudy{} luminosities yield line ratios on the order of 0.02. This is understandable as we have made no attempt to calibrate the ISM gas metallicites predicted by \galacticus{} and so our \emline{NII} luminosities are most likely being underestimated. Calibration of the galaxy metallicites will be carried out in future work and so in this work we will simply use a correction factor to correct the $\Halpha$ luminosities from \galacticus{}.

For this work we choose to assign an $\emline{NII}/\Halpha$ line ratio by cross-matching the \galacticus{} galaxies with the SDSS sample from \citet{Masters16}, who demonstrate that there exists a tight fundamental plane-like relation between the $\emline{NII}/\Halpha$ ratio, the $\emline{OIII}/{\rm H\beta}$ line ratio and the surface star formation rate of the galaxy as a proxy for sSFR, at fixed stellar mass \citep[see also][]{Faisst16a,Faisst16c}.

In Fig.~\ref{fig:sdss_vs_galacticus_line_ratios} we plot the median $\emline{NII}/\Halpha$ ratio as a function of position in the stellar mass vs. specific star formation rate (sSFR) plane. The left-hand panel shows the distribution for the SDSS galaxies from the \cite{Masters16} catalogue with signal-to-noise ratio greater than 5 in strong nebular emission lines. We use the SDSS galaxies to assign a value for the $\emline{NII}/\Halpha$ to each of the \galacticus{} galaxies by, for each \galacticus{} galaxy, identifying the nearest 5 neighbours in stellar mass vs. sSFR space from the SDSS catalogue and computing the median $\emline{NII}/\Halpha$ ratio of these 5 galaxies\footnote{Note that we have assumed a \citet{Chabrier03} IMF, whilst \citet{Masters16} originally assumed a \citet{Kroupa01} IMF. The similarity between these two IMFs means that negligible correction needs to be applied to the stellar mass to convert between them.}. This median value is then assigned to the \galacticus{} galaxy. Even though the Euclid and WFIRST missions will probe much higher redshifts beyond the range of SDSS we can use the SDSS in this approach as \citet{Masters16} argue that the fundamental plane-like relation is only weakly dependent on redshift and that, instead, the stellar mass is the dominant influence. 

The distribution of $\emline{NII}/\Halpha$ ratio in stellar mass vs. sSFR space for \galacticus{} galaxies is shown in the right-hand panel of Fig.~\ref{fig:sdss_vs_galacticus_line_ratios}. For the \galacticus{} galaxies we reassuringly see the same trend of increasing $\emline{NII}/\Halpha$ with increasing stellar mass and decreasing sSFR, at least for the regions of stellar mass vs. sSFR space that are well sampled by the SDSS galaxies. At the extremes of the distribution, such as for galaxies with stellar masses approximately $10^8\Msol$ or $10^{12}\Msol$, we are sampling a smaller number of SDSS galaxies and so as a result the distribution of $\emline{NII}/\Halpha$ for the \galacticus{} galaxies is understandably less smooth and shows more sub-structure.

Using this approach we can then define the $\Halpha+\emline{NII}$ blended flux, $f_{\Halpha+\emline{NII}}$, as
\begin{equation}
f_{\Halpha+\emline{NII}} = f_{\Halpha}\left (1+\frac{\emline{NII}}{\Halpha}\right ),
\end{equation}
where $f_{\Halpha}$ is the $\Halpha$ flux from \galacticus{} and $\emline{NII}/\Halpha$ is the line ratio sampled from the SDSS catalogue.

Further examination of the fundamental plane-like relation reported by \citet{Masters16} and how it can be used to predict the $\emline{NII}/\Halpha$ ratio for a stellar mass selected sample is investigated in \citet{Faisst17}.

\subsection{Examination of dust attenuation}
\label{sec:dust_calibration}

\begin{figure*}
  \centering
  \includegraphics[width=0.98\textwidth]{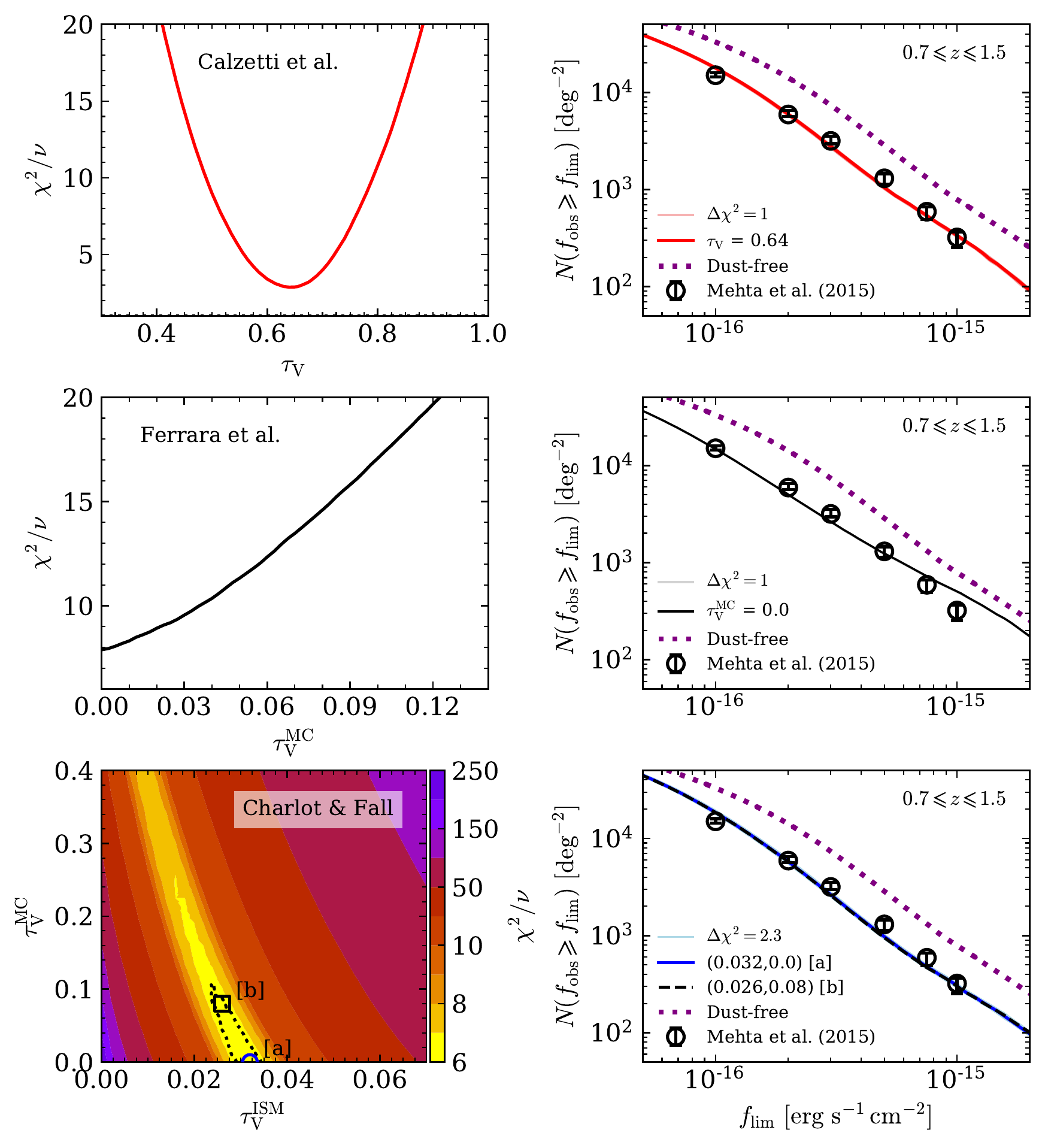}
  \caption{Calibration of the parameters for the three dust methods applied to \galacticus{}: \protect\citet[][top row]{Calzetti00}, \protect\citet[][middle row]{Ferrara99} and \protect\citet[][bottom row]{Charlot00}. For each dust method we compute the \galacticus{} number counts at a grid of points spanning the optical depth parameter space: a 1-dimensional space for the \protect\citeauthor{Calzetti00} and \protect\citeauthor{Ferrara99} methods; and a 2-dimensional space for the \protect\citeauthor{Charlot00} method. At each point we compare the \galacticus{} counts to the observed counts from \citet[][shown by the open circles in the right-hand column]{Mehta15} and compute $\chi^2/\nu$, the $\chi^2$ statistic divided by the number of degrees of freedom, $\nu$. The left-hand column shows $\chi^2/\nu$ as a function of the dust parameters.  For the \protect\citeauthor{Charlot00} method, the minimum occurs at $\left (\tau_{\rm V}^{\rm ISM},\tau_{\rm V}^{\rm MC}\right)=(0.032,0.0)$ (point [a]), whilst the dotted contour indicates the parameter space for which $\Delta\chi^2<2.3$. In the right-hand column the thick solid lines show the cumulative counts obtained when adopting the values for the optical depths for which $\chi^2/\nu$ is minimised, as reported in Table~\ref{tab:chiSquaredParameters}. The lighter solid lines show the counts for the grid points for which $\Delta\chi^2<1$, or $\Delta\chi^2<2.3$ for the \protect\citet{Charlot00} method, which correspond to the parameter uncertainties reported in Table~\ref{tab:chiSquaredParameters}. For the \protect\citeauthor{Charlot00} method the counts at $\left (\tau_{\rm V}^{\rm ISM},\tau_{\rm V}^{\rm MC}\right)=(0.026,0.08)$ (point [b]) are additionally shown.}
  \label{fig:dust_calibration}
\end{figure*}

\subsubsection{Chi-squared minimisation}
\label{sec:chi2_calibration}

For each of the three dust methods we construct a grid spanning the optical depth parameter space. For the \citet{Ferrara99} and \citet{Calzetti00} methods this parameter space is one dimensional. In the \citet{Ferrara99} method the parameter being varied is the optical depth of molecular clouds, $\tau^{\rm MC}_{\rm V}$, which is computed using Eq.~\ref{eq:opticalDepthClouds} from the \citet{Charlot00} method. The \citet{Calzetti00} method assumes a single optical depth, $\tau_{\rm V}$, for the whole galaxy. We allow this optical depth to vary, but we assume $R_{\rm V}$ to be fixed at the value $R_{\rm V}=4.05$. The \citet{Charlot00} method has a two dimensional parameter space, as we allow the normalisation for the optical depth of the ISM, $\tau^{\rm ISM}_{\rm V}$, and the normalisation for the optical depth of molecular clouds, $\tau^{\rm MC}_{\rm V}$, to vary. At each grid point in the parameter spaces we apply dust attenuation to the \galacticus{} fluxes, using the corresponding dust parameter values, and compute the cumulative number counts. To quantify the comparison to the \citet{Mehta15} observed counts we compute the $\chi^2$ goodness-of-fit statistic, defined as $\chi^2=\sum (O-E)^2/\sigma^2_O$, where $O$ is the observed counts from \citet{Mehta15}, $E$ is the expected counts from \galacticus{} and $\sigma_O$ is the uncertainty on the \citet{Mehta15} counts. Since the \citet{Mehta15} counts have asymmetric positive and negative uncertainties we follow the approach of \citet{Wang02} and define $\sigma_O=\sigma_{-}$ when $O\geqslant E$ and $\sigma_O=\sigma_{+}$ otherwise. The reduced $\chi^2$ is $\chi^2/{\nu}$, where $\nu=N_{\rm data}-N_{\rm par}$ is the number of degrees of freedom. We have $N_{\rm data}=6$ (\citealt{Mehta15} data points), and $N_{\rm par}=1$ for the \citet{Calzetti00} and \citet{Ferrara99} methods, and $N_{\rm par}=2$ for the \citet{Charlot00} method.

In Table~\ref{tab:chiSquaredParameters} we report the minimum values for the $\chi^2/\nu$ statistic along with the optical depth parameter values that yield these minimum values. Uncertainties on the parameters, which correspond to $\pm1\sigma$, are estimated by identifying, for each parameter, the range of values for which $\Delta\chi^2=1$ for the \citet{Ferrara99} and \citet{Calzetti00} methods, and $\Delta\chi^2=2.3$ for the \citet{Charlot00} method. The values for $\chi^2/\nu$ as a function of the optical depth parameters are shown in the left-hand column of Fig.~\ref{fig:dust_calibration}, whilst in the right-hand column we show the cumulative counts for the optical depth values reported in Table~\ref{tab:chiSquaredParameters}. In the right-hand column of Fig.~\ref{fig:dust_calibration} the faint solid lines show the counts for all of the optical depth parameters that yield counts for which $\Delta\chi^2=1$, or $\Delta\chi^2=2.3$ for the \citet{Charlot00} method, which corresponds to $\pm1\sigma$ uncertainty.

Considering the the left-hand column of Fig.~\ref{fig:dust_calibration} we see that all three of the dust methods show clear minima in their $\chi^2/\nu$ distributions, though of the three methods the \citet{Calzetti00} method is the best fit, as can be seen from Table~\ref{tab:chiSquaredParameters} where the \citet{Calzetti00} method has a smaller value for $\chi^2_{\rm min}/\nu$ than the other methods. With a minimum $\chi^2_{\rm min}/\nu$ value of approximately 8 the \citet{Ferrara99} method yields the poorest fit to the observed counts, possibly due it over-predicting the counts at bright fluxes around $1\times10^{-15}\ergPerSecondPerCM$. The poorness of fit is perhaps not surprising given that this method is constrained by the dust attenuation of the ISM being computed directly from the galaxy properties, with no parametrisation. For the \citet{Ferrara99} method $\chi^2/{\nu}$ is minimised when $\tau^{\rm MC}_{\rm V}=0$, implying that the best fit to the WISP counts requires no, or very little, nebular dust attenuation. However, the uncertainty on $\tau^{\rm MC}_{\rm V}$ suggests that weak nebular attenuation is permissible, though increasing the optical depth leads to \galacticus{} under-predicting the counts. Combined with the large value for $\chi^2_{\rm min}/{\nu}$, this could also suggest that the ISM attenuation set by the galaxy properties is too strong and that further calibration of \galacticus{} is required in the future.

\begin{table}
\centering
\caption{Results of $\chi^2$ minimisation procedure to compare \galacticus{} counts to observed counts from \citet{Mehta15}. Shown for each dust method is the minimum value for the reduced $\chi^2$ statistic, with $\nu=5$ degrees of freedom for the \citet{Ferrara99} and \citet{Calzetti00} methods and $\nu=4$ degrees of freedom for the \citet{Charlot00} method. The optical depth parameters that yield this minimum value for $\chi^2/\nu$ are reported along with an estimate of their uncertainties. The uncertainties correspond to $\pm1\sigma$ and are estimated by identifying for each parameter the range of values for which $\Delta\chi^2=1$ for the \citet{Ferrara99} and \citet{Calzetti00} methods, and $\Delta\chi^2=2.3$ for the \citet{Charlot00} method.}
\begin{tabular}{|c|c|l|}
\hline
Dust Method&$\chi^2_{\rm min}/\nu$& Parameter Values\\
\hline\hline
\input{chiSquaredDustParameters.tex}
\hline
\end{tabular}
\label{tab:chiSquaredParameters}
\end{table}

\begin{figure}
  \centering
  \includegraphics[width=0.48\textwidth]{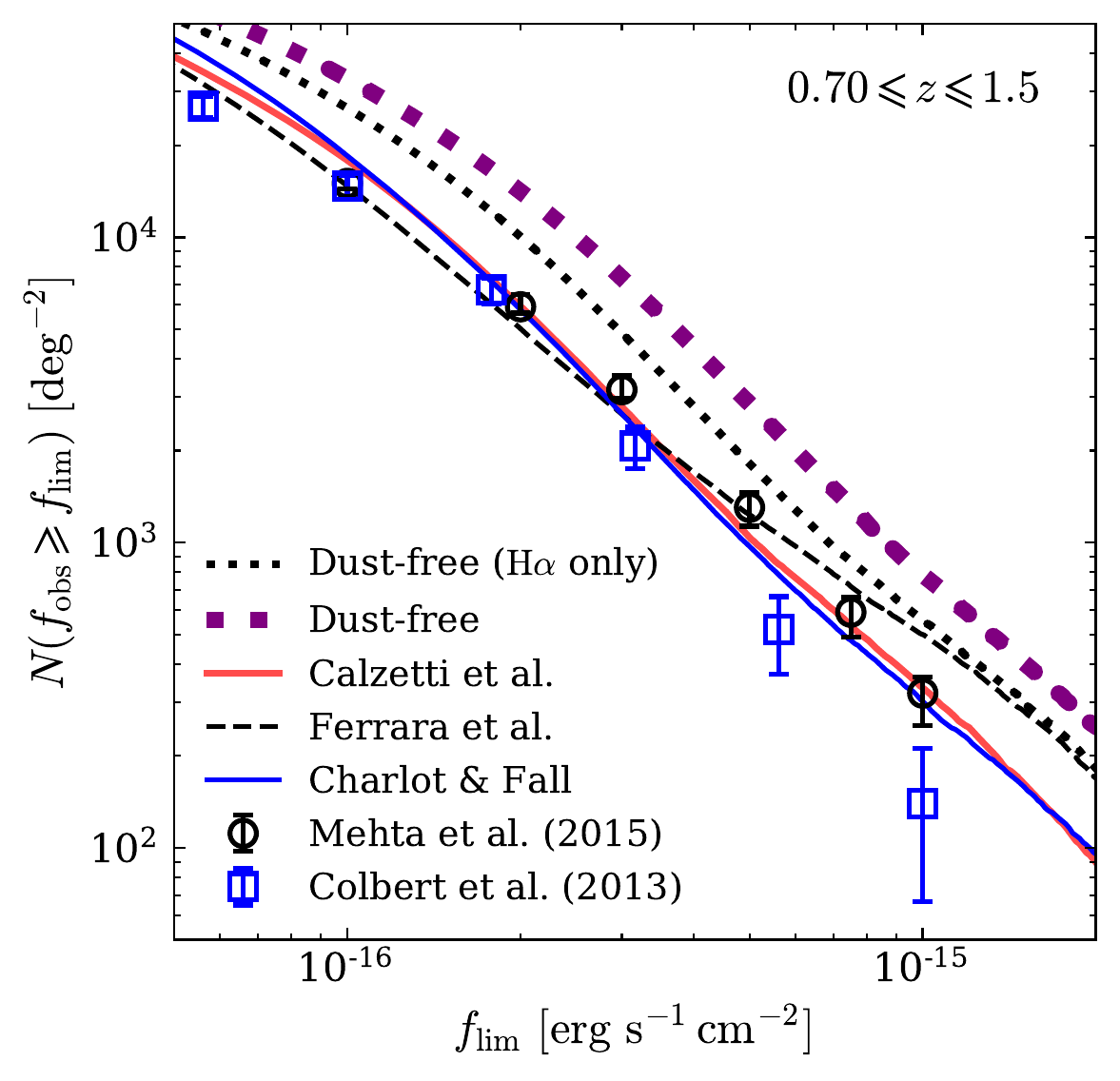}
  \caption{The cumulative flux counts for the redshift range $0.7\leqslant z\leqslant 1.5$ from the \galacticus{} lightcone mock catalogue. The various lines show the predictions for the $\Halpha$-only fluxes and the $\Halpha+{\rm \emline{NII}}$ blended fluxes. The observed counts reported by \protect\citet{Colbert13} and \protect\citet{Mehta15} are shown by the symbols. For the three dust methods we assume values for the optical depths as stated in Table~\ref{tab:chiSquaredParameters}.}
  \label{fig:halpha_flux_counts}
\end{figure}

For the \citet{Charlot00} method the minimum reduced $\chi^2$ value is approximately $\chi^2_{min}/\nu \simeq 6$ and is found at approximately $\left (\tau^{\rm ISM}_{\rm V},\tau^{\rm MC}_{\rm V} \right ) \simeq \left ( 0.032, 0.0\right )$. As with the \citet{Ferrara99} method, this again suggests a requirement for no or very weak nebular dust attenuation. However a clear degeneracy between the optical depth parameters is clearly evident, with the shape of this degeneracy suggesting that the ISM optical depth is the more sensitive parameter when attempting to match the WISP counts. As such, parameter combinations with $\tau^{\rm MC}_{\rm V}\lesssim 0.1$ are identified as having $\Delta\chi^2=2.3$, as indicated by the dotted contour in the bottom left-hand panel of Fig.~\ref{fig:dust_calibration}. Parameter combinations along this degeneracy yield as good a match to the WISP counts, as seen in the bottom right-hand panel of Fig.~\ref{fig:dust_calibration} where we additionally show the counts for the parameter combination $\left (\tau^{\rm ISM}_{\rm V},\tau^{\rm MC}_{\rm V} \right ) \simeq \left ( 0.026, 0.08\right )$. The allowed values for the $\tau^{\rm ISM}_{\rm V}$ and $\tau^{\rm MC}_{\rm V}$ parameters are much smaller than the values originally determined by \citet{Charlot00}, although we note that we are calibrating against a different subset of the galaxy population at a much higher redshift.

In Fig.~\ref{fig:halpha_flux_counts} we show again the cumulative number counts for each of the three dust methods, obtained using the optical depth parameters reported in Table~\ref{tab:chiSquaredParameters}. In addition, for completeness, we also show the cumulative number counts for the \galacticus{} galaxies when no dust attenuation is applied, both for the $\Halpha$ fluxes and the $\Halpha+\emline{NII}$ blended fluxes. The observed counts from \citet{Colbert13} and \citet{Mehta15} are also shown. We can see that the \citet{Calzetti00} and \citet{Charlot00} methods yield very similar counts, whilst the \citet{Ferrara99} method yields slightly higher counts for fluxes brighter than approximately $5\times10^{-16}\ergPerSecondPerCM$.

In summary, by minimising the $\chi^2$ statistic we are able to identify optical depth parameters for each dust method with which we can obtain \galacticus{} counts that are broadly consistent with the WISP counts, particularly at flux limits fainter than $2\times 10^{-16}\ergPerSecondPerCM$. However, the values we obtain for $\chi^2/\nu$ suggest that further calibration or modification of \galacticus{} or the dust methods is needed in the future. At intermediate flux limits around $5\times 10^{-16}\ergPerSecondPerCM$ the slope of the counts obtained with the \citet{Calzetti00} and \citet{Charlot00} methods leads to these methods under-predicting the \citet{Mehta15} counts by up to a factor of 2, as can be seen in Fig.~\ref{fig:halpha_flux_counts}. At bright fluxes of $1\times 10^{-15}\ergPerSecondPerCM$, the \citet{Ferrara99} method yields counts that are approximately a factor of 2 higher than the WISP counts. Lightcone catalogues built from other semi-analytical models have previously under-predicted the number of galaxies and been unable to match the \citet{Mehta15} observations (we refer the reader to the discussion in \citealt{Pozzetti16}). One possible reason that \galacticus{} is able to reproduce the \citet{Mehta15} counts could be the use of \cloudy{}, which has not been previously coupled to other semi-analytical models. Alternatively, the agreement could be a result of our choice of dust attenuation methods. However, there are substantial differences between semi-analytical models, both in how they model the astrophysical processes and how they are calibrated, which would affect the predicted number counts. Therefore understanding the cause of the disagreement will require detailed comparison of \galacticus{} with other semi-analytical models \citep[e.g. see][]{Knebe15}, which we leave for future work.

\subsubsection{Comparison to WISP optical depths}

\begin{figure}
  \centering
  \includegraphics[width=0.48\textwidth]{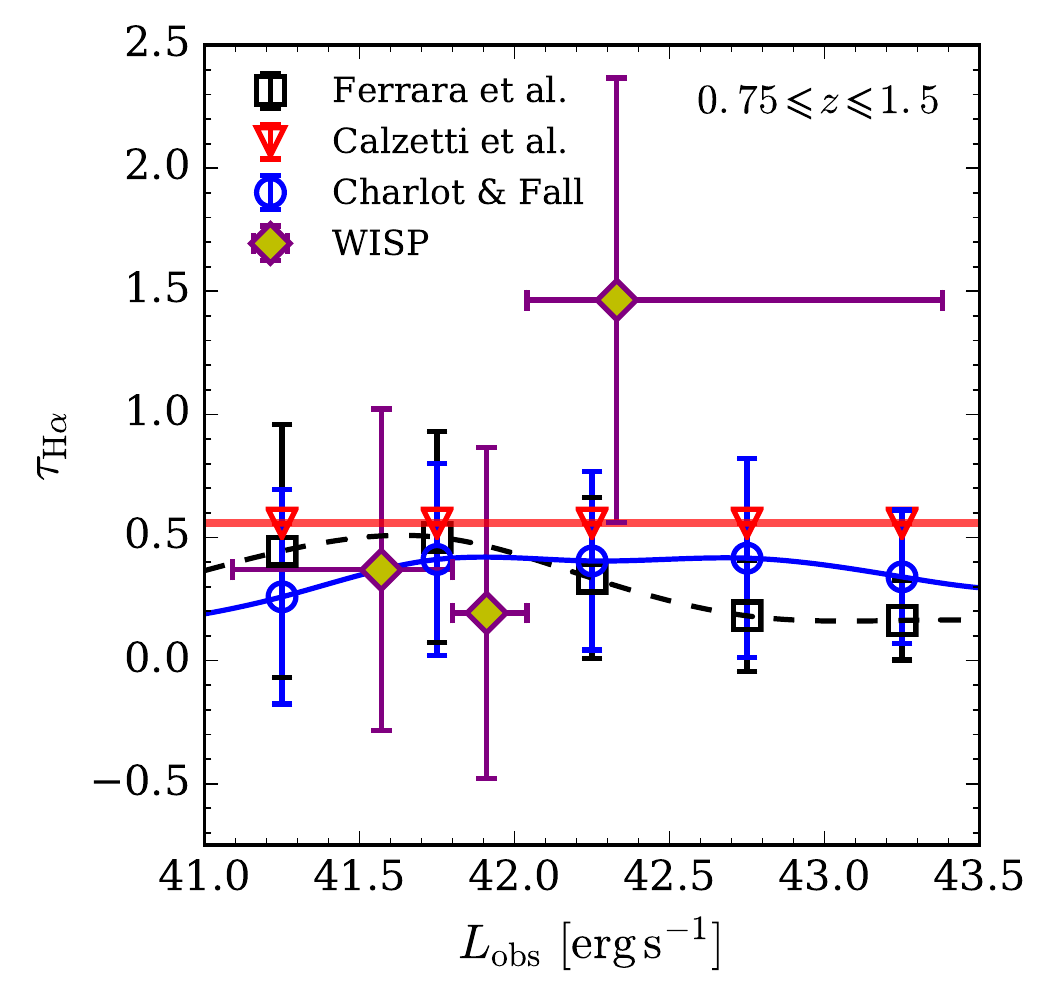}
  \caption{The optical depth at the $\Halpha$ wavelength, $\tau_{\rm \Halpha}$ for the \galacticus{} galaxies when processed with each of the three dust methods: \protect\citet{Ferrara99}, \protect\citet{Charlot00} and \protect\citet{Calzetti00}. The parameters for each method correspond to the optimal values stated in Fig.~\ref{fig:dust_calibration}. For \galacticus{} we show the mean and $\pm1\sigma$ dispersions in bins of de-blended dust-attenuated $\Halpha$ luminosity over the redshift range $0.75\leqslant z\leqslant 1.5$. For comparison we show the WISP results from \protect\citet{Dominguez13}.}
  \label{fig:optical_depth}
\end{figure}

Our examination of the dust parameters has primarily provided us with a set of optical depth parameters that will yield \galacticus{} fluxes whose counts we know will show a reasonable agreement with WISP. Matching the WISP counts is necessary for making predictions for Euclid and WFIRST. These results also tentatively suggest that weak dust attenuation is required to match the WISP counts. Several observations hint at high redshift $\Halpha$ emitting galaxies having little dust attenuation \citep[e.g.][]{Dominguez13, Masters14, Reddy15}. For each \galacticus{} galaxy we can compute the overall dust attenuation at the $\Halpha$ line wavelength according to,
\begin{equation}
\tau_{\Halpha} = -\ln\left (\frac{L^{\rm att}_{\rm \Halpha}}{L^0_{\rm \Halpha}}\right ),
\end{equation}
where $\tau_{\Halpha}$ is the optical depth at the $\Halpha$ wavelength, $L^{\rm att}_{\Halpha}$ is the dust attenuated $\Halpha$ luminosity and $L^0_{\rm \Halpha}$ is the dust-free, intrinsic $\Halpha$ luminosity of the galaxy. In Fig.~\ref{fig:optical_depth} we plot for each dust method the mean value for $\tau_{\Halpha}$ (and $1\sigma$ uncertainty) as a function of de-blended dust-attenuated $\Halpha$ luminosity. In each case we assume the attenuation parameters that minimise $\chi^2/{\nu}$, as reported in Table~\ref{tab:chiSquaredParameters}. For comparison we show the estimates from the WISP analysis of \citet{Dominguez13}. Note that attenuation computed by the \citet{Calzetti00} method depends only on wavelength and so the attenuation is identical for each galaxy, leading to a constant mean optical depth with zero scatter.

\begin{figure*}
  \centering
  \includegraphics[width=0.98\textwidth]{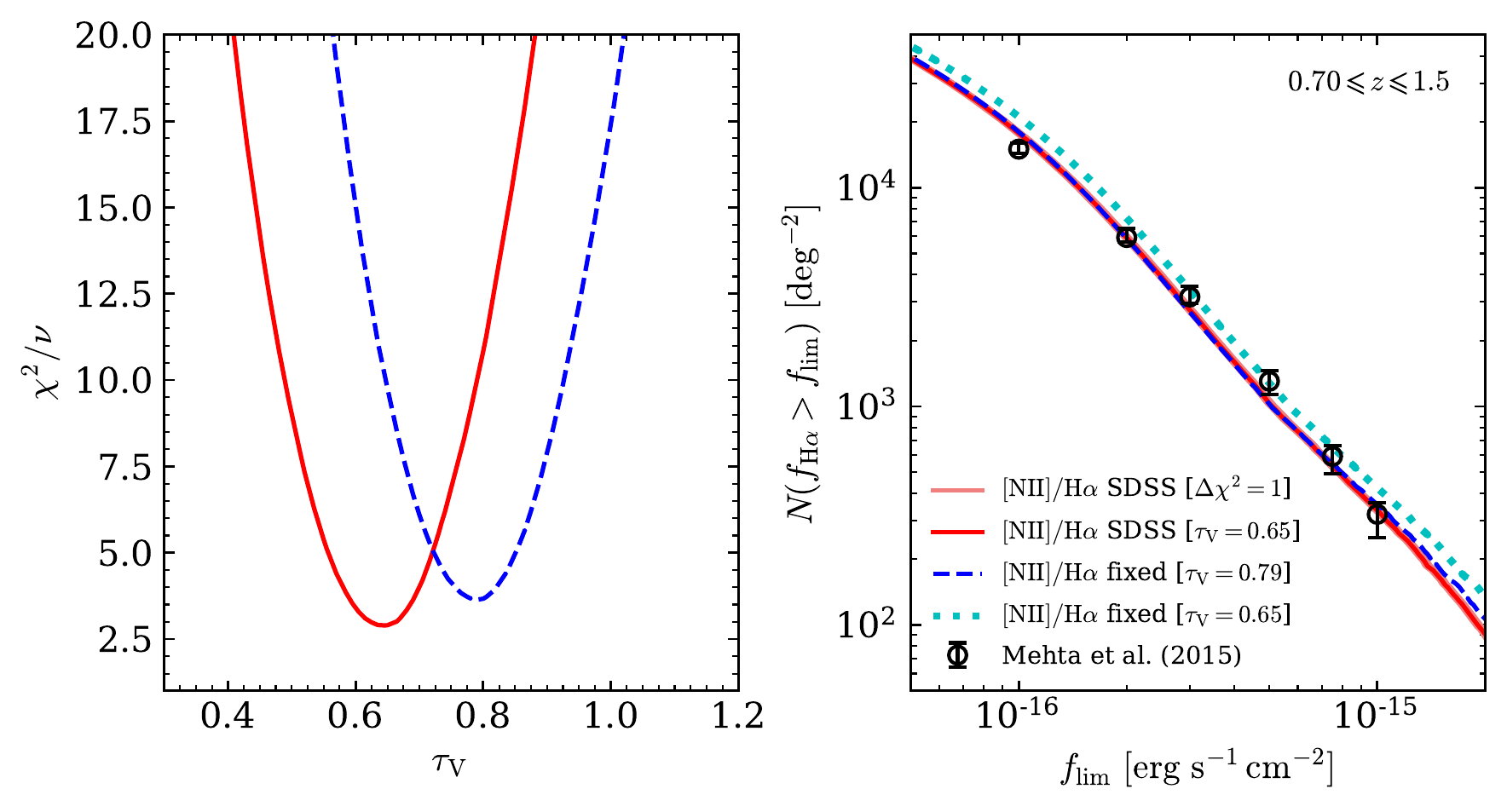}
  \caption{Calibration of the optical depth used in the \protect\citet{Calzetti00} dust method in the case where \emline{NII} contamination is fixed at 29 per cent and the case where \emline{NII} contamination is set by assigning the ${\rm \emline{NII}}/\Halpha$ line ratio using the nearest neighbours in the \protect\citet{Masters16} SDSS catalogue. The left-hand panel shows the reduced $\chi^2$ statistic, $\chi^2/{\nu}$, for these two cases as a function of optical depth. The right-hand panel shows the cumulative counts from dust attenuated \galacticus{} fluxes when adopting the optical depth that minimises $\chi^2/{\nu}$. The faint red lines show the counts obtained when assuming optical depths within $\Delta\chi^2=1$ (which corresponds to $\pm1\sigma$) for the case in which the ${\rm \emline{NII}}/\Halpha$ line ratio is assigned using the SDSS galaxies. The dotted cyan line shows the counts obtained when adopting a fixed ${\rm \emline{NII}}/\Halpha$ line ratio but using the optical depth obtained when assuming line ratios from SDSS.}
  \label{fig:counts_sdss_line_ratio}
\end{figure*}

We can see that each of the three dust methods yield optical depths that are consistent within error with the WISP measurements. For luminosities fainter than $10^{42.5}\ergPerSecond$ all three methods are consistent within error with one another, whilst for the brightest luminosities there is growing disagreement between the \citet{Ferrara99} and \citet{Calzetti00} methods. From their measurements, \cite{Dominguez13} report an increase in the optical depth with increasing observed luminosity. A similar trend is also reported by \citet{Hopkins01} and \citet{Sobral12}. However, the \galacticus{} optical depths for all three dust methods are consistent with no correlation between optical depth and observed luminosity. Assuming that this trend is real and not, for example, due to a selection bias, then the lack of such a trend in the \galacticus{} predictions could indicate that further calibration of the model is required. For instance, further calibration of the galaxy metallicities of bright galaxies may boost the optical depths predicted by the \citet{Ferrara99} dust method. For the \citet{Charlot00} method, the lack of trend could suggest the need for further parametrisation. We note however that due to the size of uncertainties of their measurements, the WISP optical depths from \cite{Dominguez13} could also be consistent with a flat relation and no change in optical depth with luminosity.

\subsection{Examination of \emline{NII} contamination}
\label{sec:nii_contamination}

As mentioned in Section~\ref{sec:modellingNII}, \citet{Colbert13} and \citet{Mehta15} in their  analyses adopt a fixed \emline{NII} contamination to match the $\emline{NII}/\Halpha$ line ratios presented by \citet{Villar08} and \citet{Cowie11} at $z\sim 0$. Although \citeauthor{Villar08} and \citet{Cowie11} find the $\emline{NII}/\Halpha$ line ratio to decrease with increasing $\Halpha$ equivalent width, \citet{Mehta15} argue that the line ratios measured by these authors are approximately constant for equivalent widths below $200\angstrom$. \citet{Mehta15} state that by assuming a fixed contamination, their \emline{NII} contamination will be over-estimated for only 10 per cent of their catalogue. It is worth briefly examining the impact that fixing the \emline{NII} contamination has on the dust parameters.

In the left-hand panel of Fig.~\ref{fig:counts_sdss_line_ratio} we show again the reduced $\chi^2$ statistic as a function of the optical depth for the \citet{Calzetti00} dust method when assuming $\emline{NII}/\Halpha$ ratios as assigned from the SDSS catalogue. We additionally show the reduced $\chi^2$ statistic when assuming the \emline{NII} contamination is fixed at 29 per cent, i.e.
\begin{equation}
f_{\Halpha+\emline{NII}} = \frac{f_{\Halpha}}{0.71} = f_{\Halpha}\left (1+\frac{\emline{NII}}{\Halpha}\right ).
\end{equation}
By re-arranging this equation we can see that a fixed \emline{NII} contamination of 29 per cent assumed in the WISP analyses corresponds to $\emline{NII}/\Halpha\sim0.41$. The right-hand panel of Fig.~\ref{fig:counts_sdss_line_ratio} shows, for both cases, the cumulative number counts obtained when adopting the optical depth that minimises the reduced $\chi^2$ statistic. We can see that when adopting a fixed $\emline{NII}/\Halpha$ line ratio, the value for the optical depth that minimises $\chi^2/{\nu}$ becomes larger than when adopting the $\emline{NII}/\Halpha$ line ratio from SDSS. The difference in the optical depths is due to a degeneracy between the strength of the $\emline{NII}/\Halpha$ ratio and the strength of the dust attenuation that is required in order for \galacticus{} to match the observed counts from WISP. If we consider, in Fig.~\ref{fig:halpha_flux_counts}, the \galacticus{} number counts from the dust-free, $\Halpha$ fluxes (as shown by the thin dotted line), then we can see that when we introduce \emline{NII} contamination we make the fluxes brighter and boost the number counts higher (as shown by the thick dotted line). Applying dust attenuation to the fluxes is required to bring the counts back down into agreement with the WISP counts. However, for the larger $\emline{NII}/\Halpha$ ratio of $\sim0.41$, the increase in the brightness of the $\Halpha$ fluxes is larger and so the boost in the $\Halpha$ counts is larger. As a result, stronger dust attenuation is required to bring the dust-free $\Halpha+\emline{NII}$ counts down into agreement with WISP. Obviously, if we were to assume a fixed optical depth (i.e. assume a fixed dust attenuation) then adopting the fixed $\emline{NII}/\Halpha$ ratio will lead to the \galacticus{} counts being incorrectly boosted above the counts from WISP, as shown by the thick dotted line in the right-hand panel of Fig.~\ref{fig:counts_sdss_line_ratio}.

A model for the redshift evolution of the $\emline{NII}/\Halpha$ ratio as a function of stellar mass is presented in \citet{Faisst17}.

\begin{figure}
  \centering
  \includegraphics[width=0.48\textwidth]{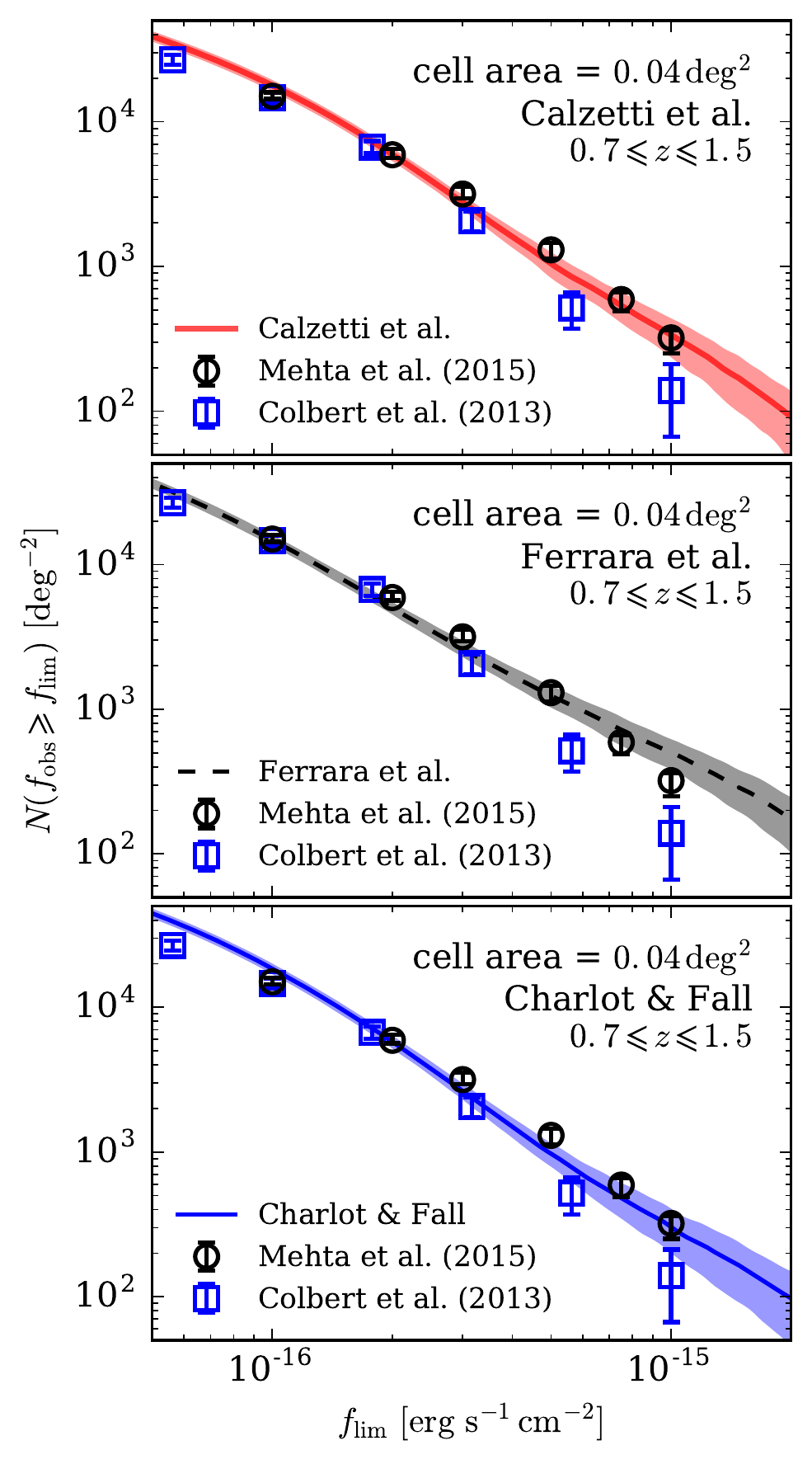}
  \caption{Distribution of cumulative $\Halpha$+\emline{NII} flux counts from \galacticus{} when the lightcone footprint is divided up into 100 cells, each with an individual area of 0.04 square degrees. The three panels show the predictions for each of the dust methods: \protect\citeauthor{Calzetti00} (top panel), \protect\citeauthor{Charlot00} (middle panel) and \protect\citeauthor{Ferrara99} (bottom panel). The lines correspond to the mean counts whilst the shaded regions correspond to the 1$\sigma$ uncertainty. For each dust method the optical depth parameters correspond to those that minimise the reduced chi-squared statistic, as reported in Table~\ref{tab:chiSquaredParameters}.}
  \label{fig:halpha_flux_counts_cosmicVariance}
\end{figure}

\begin{figure}
  \centering
  \includegraphics[width=0.48\textwidth]{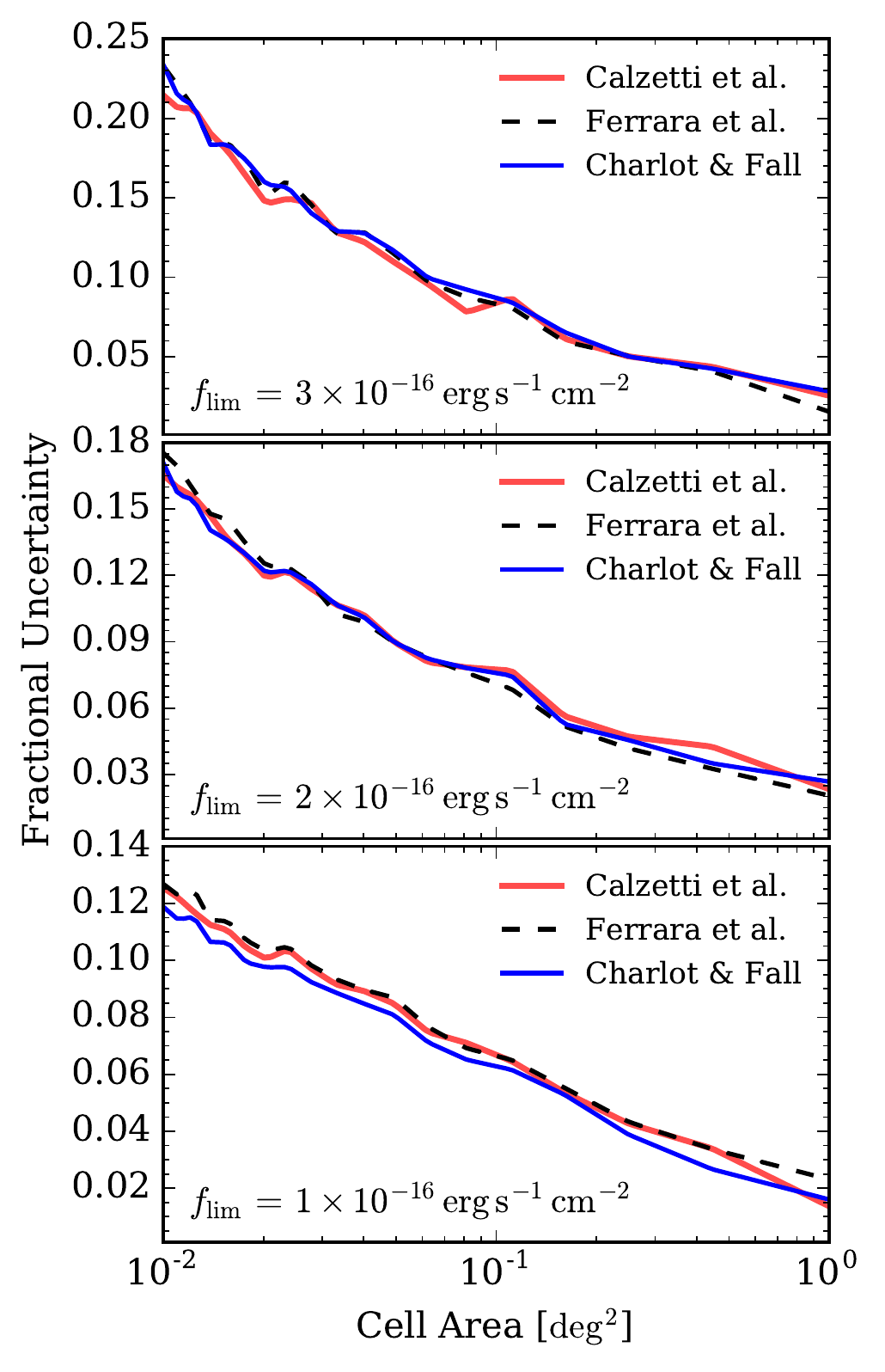}
  \caption{Fractional uncertainty in the \galacticus{} cumulative $\Halpha$+\emline{NII} flux counts due to cosmic variance as a function of cell size in square degrees for three different flux limits, $f_{\rm lim}$, as labelled in the bottom left-hand corner of each panel. The lightcone footprint is split into $N\times N$ cells and the counts computed for each cell. For each flux limit, we compute the mean and standard deviation in the counts over the cells. The fractional uncertainty is defined as the ratio of the standard deviation to the mean counts.}
  \label{fig:cosmicVarianceVsArea}
\end{figure}

\begin{figure*}
  \centering
  \includegraphics[width=0.98\textwidth]{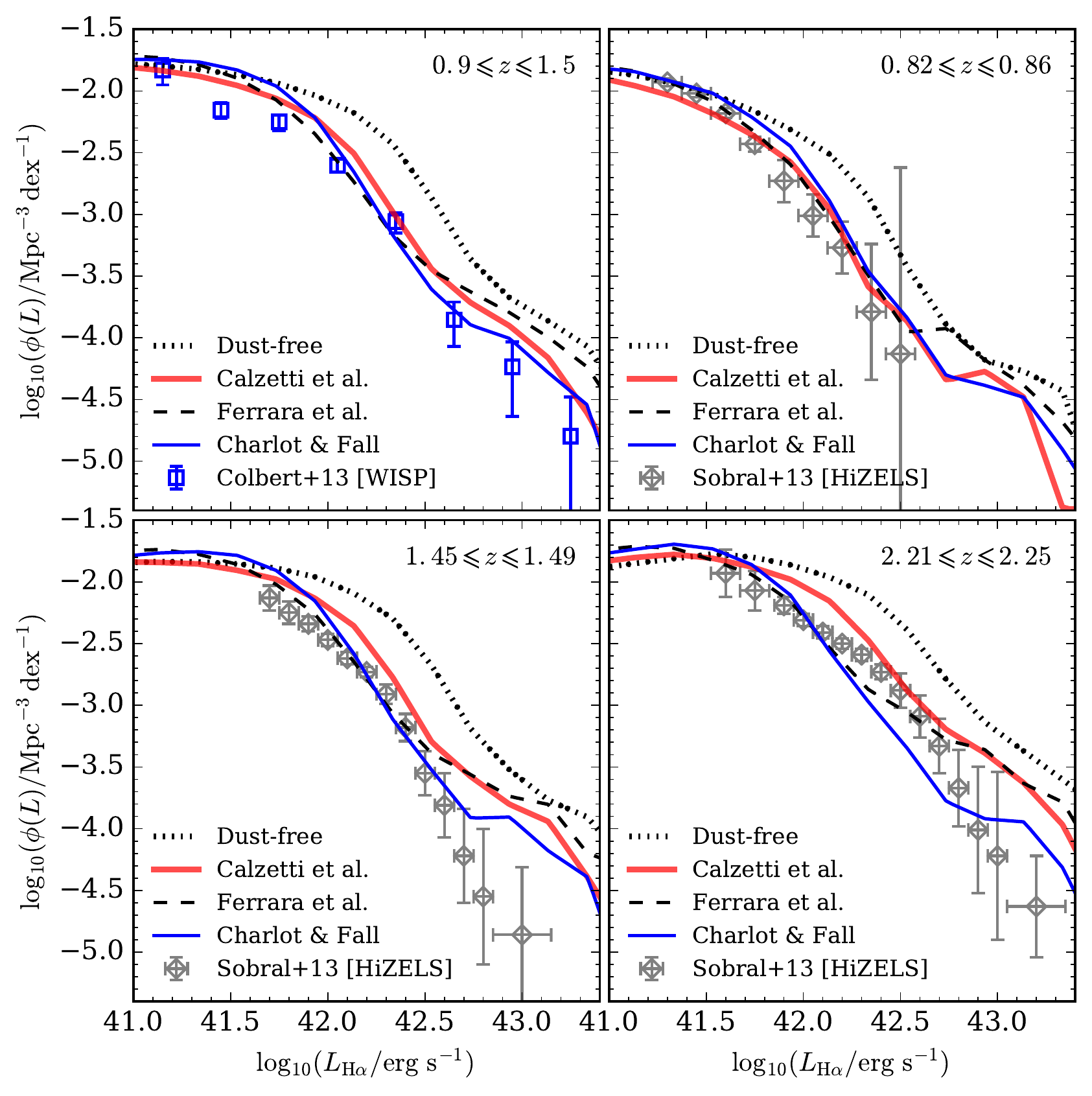}
  \caption{\galacticus{} predictions for the dust-attenuated rest-frame $\Halpha$ luminosity function estimated for a broad redshift bin in the top left-hand panel, and for three narrow redshift bins, centred at $z=0.84$, $z=1.47$ or $z=2.23$, as shown in the remaining three panels. The redshift bin is indicated in the top right-hand corner of each panel. Observational estimates from \protect\citet{Colbert13} and \protect\citet{Sobral13} are shown by the data points in the panels for the appropriate redshift interval. The various lines show the \galacticus{} predictions, with the line colour and style corresponding to the same dust method as in Fig.~\ref{fig:halpha_flux_counts}. For each dust method we adopt the optical depth parameters that minimise the reduced chi-squared statistic, as reported in Table~\ref{tab:chiSquaredParameters}. In addition, in each panel we show the \galacticus{} luminosity function for the dust-free fluxes.}
  \label{fig:halpha_lf}
\end{figure*}

\subsection{Examination of cosmic variance}
\label{sec:cosmic_variance}

Another source of uncertainty on the number counts is cosmic variance. Although the sky area of the \galacticus{} lightcone is approximately an order of magnitude larger then the total combined area of all the WISP fields, it is worth briefly examining how cosmic variance is impacting the counts. We make an estimate of the impact of cosmic variance by splitting the lightcone sky area up into $N\times N$ cells and computing the cumulative number counts in each cell. In Fig.~\ref{fig:halpha_flux_counts_cosmicVariance} we show the mean and standard deviation for the counts computed using $10 \times 10$ cells, each with an area of $0.04\,{\rm deg}^{2}$. Note that in their analysis \citet{Mehta15} worked with a total area of approximately $0.051\,{\rm deg}^{2}$. We estimate that for a flux limit of $1\times 10^{-16}\ergPerSecondPerCM$ there is approximately 8 per cent variation in the total counts. For a flux limit of $3\times 10^{-16}\ergPerSecondPerCM$ this grows to approximately 12 per cent, whilst for bright flux limits of approximately $1\times 10^{-15}\ergPerSecondPerCM$ we see approximately a factor of 30 per cent variation.

In Fig.~\ref{fig:cosmicVarianceVsArea} we plot the fractional uncertainty in the total counts as a function of cell area for three different flux limits: $3\times 10^{-16}\ergPerSecondPerCM$, $2\times 10^{-16}\ergPerSecondPerCM$ and $1\times 10^{-16}\ergPerSecondPerCM$. The fractional uncertainty is defined as the ratio of standard deviation to the mean counts at that flux limit. For a sky area of 1 square degree we see variations of approximately 2--3 per cent in the counts. Since the uncertainty is decreasing with increasing area, for an area of 4 square degrees we expect the variation in the counts due to cosmic variance to be even smaller, on the order of 1 per cent. We do not examine cosmic variance further in this work but note that the full impact of cosmic variance will need to be properly examined in future work when larger area lightcones become available.

\section{Luminosity Functions}
\label{sec:luminosity_functions}

As a further examination of the \galacticus{} model and our chosen dust attenuation methods, we consider the $\Halpha{}$ galaxy luminosity function, which describes the number density of $\Halpha$ emitting galaxies as a function of their luminosity. In Fig.~\ref{fig:halpha_lf} we plot the \galacticus{} predictions for the dust-attenuated, rest-frame $\Halpha$ luminosity function for several redshift intervals chosen to match intervals used in observational analyses by \citet{Colbert13} and \citet{Sobral13}. For the comparisons with the observations from \citet{Sobral13} we simply take a narrow redshift interval centred on the appropriate redshift: $z=0.84$, $z=1.47$ or $z=2.23$. The \galacticus{} luminosity function is computed from the lightcone prior to application of any selection criteria and so we can simply count all the galaxies within the specified redshift interval without needing to apply any further volume re-weighting (such as the in the maximum volume $V_{\rm max}$ approach). Note that all luminosity functions in Fig.~\ref{fig:halpha_lf} are corrected to a Hubble parameter of $h=0.7$.

Focussing to begin with on the top left-hand panel of Fig.~\ref{fig:halpha_lf}, which compares the \galacticus{} predictions with the luminosity function estimated by \citeauthor{Colbert13}, we see that predicted luminosity function for each of the dust methods has broadly the correct normalisation and shape consistent with the WISP observations, though predict an excess of bright galaxy counts for luminosities brighter than $10^{43}\ergPerSecond$. All three dust methods yield an excess of faint galaxies at luminosities $\sim 10^{41.5}\ergPerSecond$. 

Considering the narrow redshift ranges in the remaining three panels of Fig.~\ref{fig:halpha_lf}, we see that at the lower redshift of $z\simeq0.84$ all of dust methods yield luminosity functions that are consistent with the HiZELS estimate from \citet{Sobral13}. At the higher redshifts of $z\simeq1.47$ and $z\simeq2.23$ however, all of the dust methods yield luminosity functions that have an incorrect shape. At $z\simeq1.47$ the \galacticus{} predictions are consistent with the HiZELS estimates for luminosities between $10^{42}\ergPerSecond$ and $10^{42.5}\ergPerSecond$, but for brighter luminosities all of the methods predict an excess of bright galaxies compared to HiZELS. The worst agreement is seen at $z\simeq2.23$ where the \citet{Calzetti00} and \citet{Ferrara99} methods again predict an excess of bright galaxies compared to the observational estimates, whilst the \citet{Charlot00} method predicts a deficit of galaxies for luminosities between $10^{42}\ergPerSecond$ and $10^{42.5}\ergPerSecond$. 

At each redshift epoch the difference between the luminosity functions for the three dust methods is a direct consequence of the disagreement between the methods for the prediction for the strength of dust attenuation. This is most likely a result of the methods using different approaches to model the attenuation. We remind the reader that the \citet{Ferrara99} method is the most physical, predicting the attenuation from the galaxy properties, whilst the \citet{Calzetti00} method is the least physical, adopting an attenuation that is a function of wavelength only. Furthermore, even if the dust methods were all in agreement, the discrepancy between the model predictions and the observational estimates suggests a calibration issue, either with the dust methods or with \galacticus{} itself. For example, the discrepancy between the model predictions and the observations may indicate that the dust methods are too simplistic and require further parametrisation, perhaps as a function of redshift. Alternatively, if we were to assume that the dust attenuation is correct, then this discrepancy could, for example, indicate that the star formation rate in \galacticus{} is incorrect at high redshifts. Finally, the luminosity functions in Fig.~\ref{fig:halpha_lf} highlight the difficulty of correctly modelling the redshift evolution of dust and galaxy properties such that one can recover the statistics both at individual epochs as well as integrated statistics over a broad redshift range. Since we have previously chosen to adjust the dust attenuation to match the WISP number counts, it is therefore understandable that \galacticus{} predictions show the best agreement with the \citet{Colbert13} luminosity function over the broadest redshift range. However, we also note that narrow band surveys targeting single redshift epochs such as HiZELS can suffer from significant contamination from emission lines from galaxies at lower or higher redshifts \citep[e.g. see][]{Martin08,Henry12,Colbert13}, which may be distorting the shape of the measured HiZELS luminosity functions. However, \citet{Matthee17} have recently presented luminosity function measurements from 0.7 ${\rm deg}^2$ of HiZELS overlapping with the B{\"o}otes field which show number densities in good agreement with the \citeauthor{Colbert13} measurement.

Overall the normalisation of the luminosity function predicted from \galacticus{} is consistent with the various datasets, though at redshifts $z\gtrsim1.5$ the shapes of the \galacticus{} luminosity functions are incorrect, often leading to either a deficit of galaxies at intermediate luminosities or an excess of bright galaxies. This could be due to the adopted values for the parameters governing the strength of the dust attenuation for the various methods, or could perhaps suggest that the parametrisation requires further modification to account for evolution with redshift. Alternatively, the excess of bright galaxies and deficit of galaxies at intermediate luminosities could be due to factors such as the instantaneous star formation rate being too high in the model. Further investigation, including a thorough investigation of the model parameter space and examination of the galaxy counts in broad photometric bands, is left for future work.

\section{Survey forecasts}
\label{sec:forecasts}

\begin{figure*}
  \centering
  \includegraphics[width=0.98\textwidth]{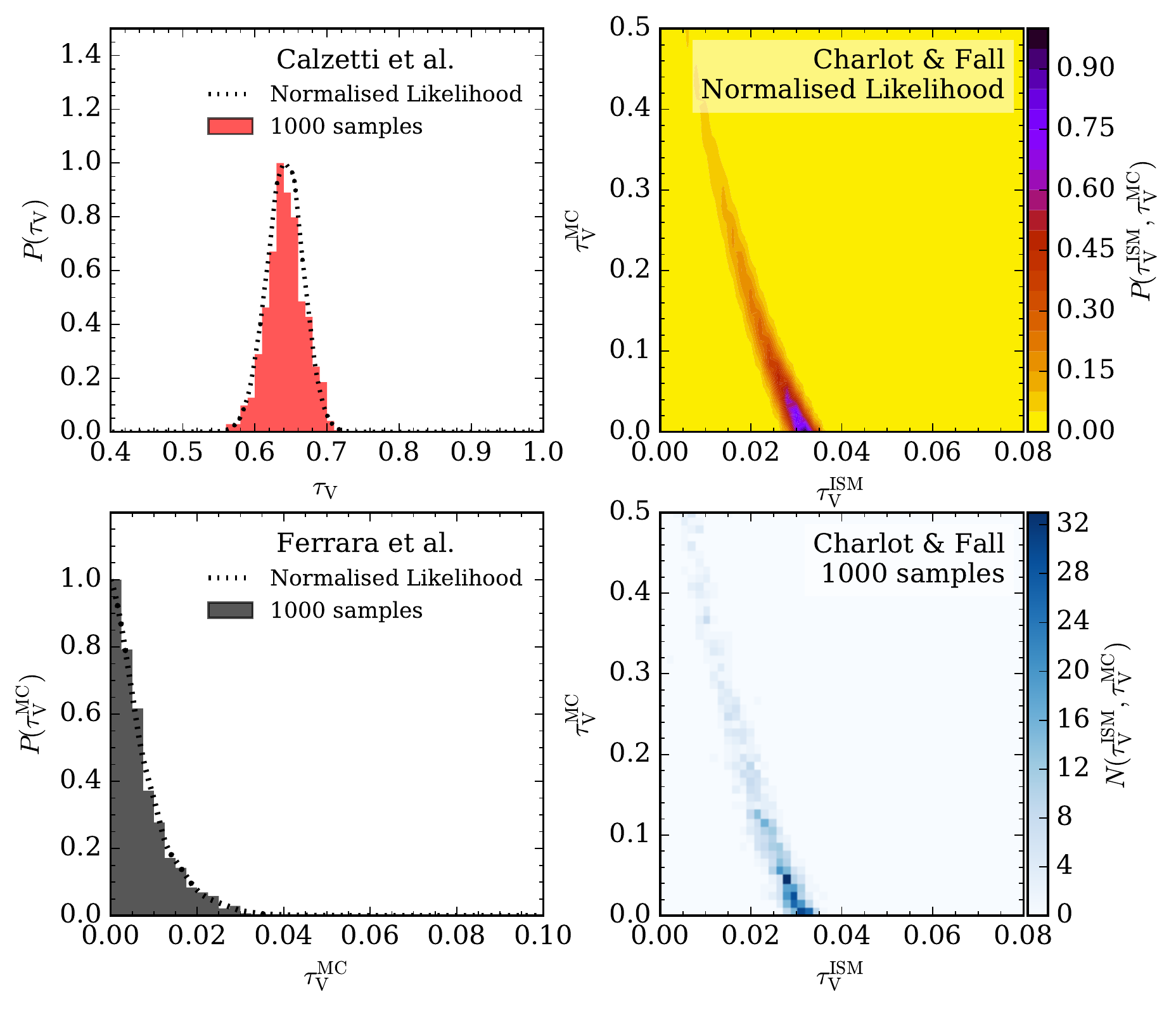}
  \caption{Normalised likelihoods and distributions for 1000 Monte-Carlo realisations of the optical depth parameters for the \protect\citet{Calzetti00} dust method (top left panel), the \protect\citet{Ferrara99} dust method (bottom left panel) and the \protect\citet{Charlot00} dust method (right-hand panels). The distribution of 1000 parameter realisations are shown by the shaded 1-dimensional, or 2-dimensional, histograms.}
  \label{fig:monteCarloSamples}
\end{figure*}

We now use our \galacticus{} lightcone to predict the $\Halpha$-selected differential and cumulative flux counts as well as the redshift counts, for survey designs similar to Euclid and WFIRST. In the case of a Euclid-like survey, we consider two possible strategies. In both instances we assume a redshift limit of $0.9\leqslant z\leqslant1.8$, however in the first instance we assume a flux limit of $3\times 10^{-16}\ergPerSecondPerCM$ and in the second instance we assume a deeper a flux limit of $2\times 10^{-16}\ergPerSecondPerCM$. These instances are consistent with two of the Euclid-like strategies presented in \citet{Pozzetti16}. For a WFIRST-like design we consider a redshift range of $1\leqslant z \leqslant 2$ and a flux limit of $1\times 10^{-16}\,\ergPerSecondPerCM$. 

\subsection{Sampling of optical depth parameters}

To account for the distribution of permissible optical depth parameters for each dust method we perform Monte-Carlo sampling of the optical depth parameters.

For each dust method we randomly sample values for the optical depth parameters, by using the appropriate $\chi^2$ from Fig.~\ref{fig:dust_calibration} to construct a likelihood, $\mathcal{L}$,
\begin{equation}
\mathcal{L} \propto \exp\left ( -\chi^2/2 \right ),
\end{equation}
with which we can randomly sample the optical depths. For each dust method we sample 1000 realisations for the optical depth parameters.
The normalised likelihoods and distributions of the 1000 realisations are shown in Fig.~\ref{fig:monteCarloSamples}.

For each dust method we therefore recompute the dust attenuation 1000 times and in each instance use the dust-attenuated $\Halpha+\emline{NII}$ blended fluxes to compute the cumulative and differential flux counts, as well as the flux-limited redshift distribution. The forecasts that we present therefore correspond to the mean and standard deviation of these 1000 realisations.

\begin{figure*}
  \centering
  \includegraphics[width=0.98\textwidth]{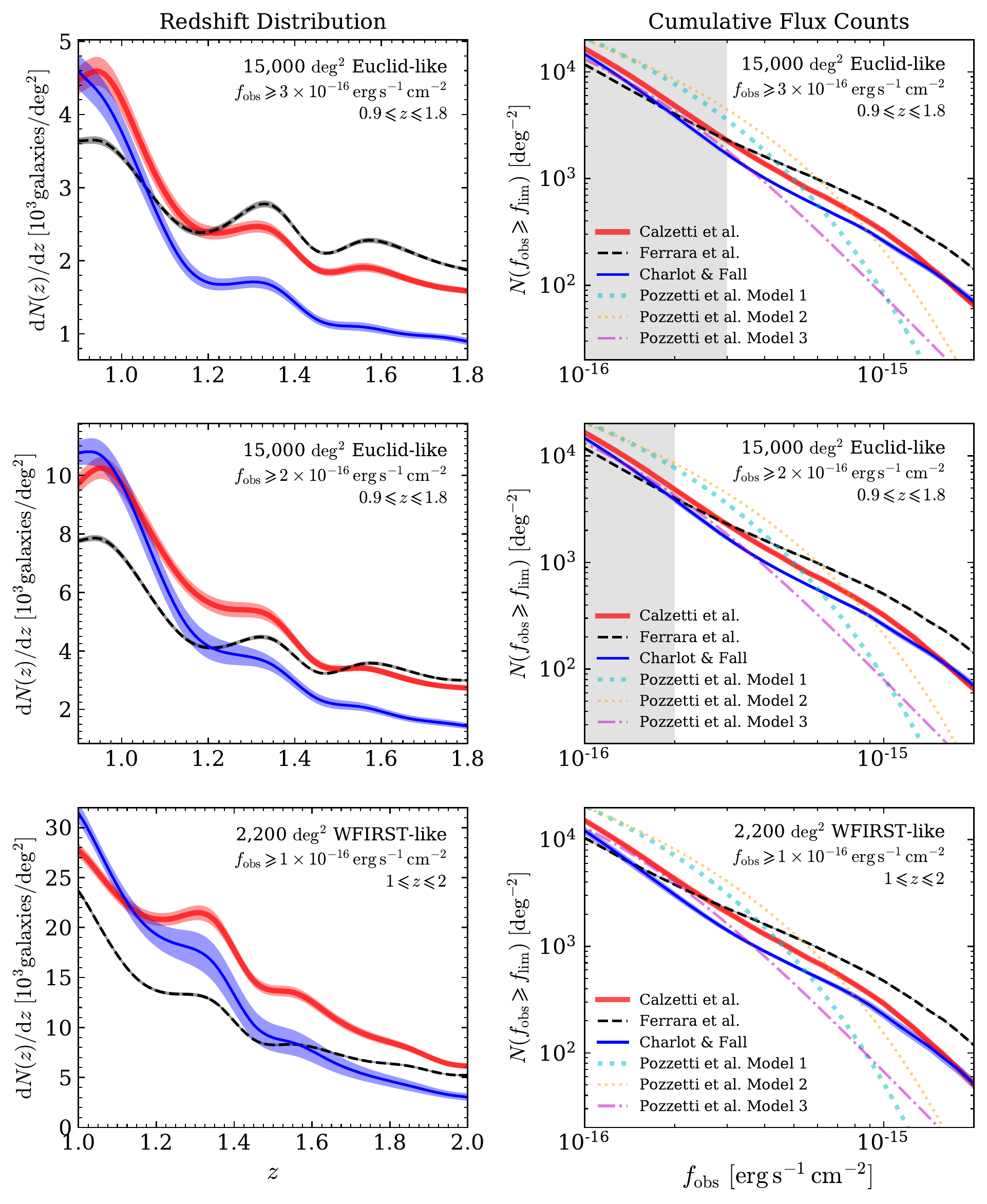}
  \caption{Predictions for the galaxy redshift distribution (left column) and cumulative flux counts (right column) for two Euclid-like survey strategies (upper and middle rows) and a WFIRST-like (bottom row) survey strategy. The corresponding flux limit and redshift range for each survey are shown in the top right-hand corner of each panel. In each panel the various lines represent the mean counts over 1000 Monte Carlo realisations, whilst the shaded regions show the $1\sigma$ uncertainty. In addition, for reference, dotted and dashed lines show the predictions for the three empirical models from \protect\citet{Pozzetti16}, corrected to show de-blended counts values by assuming a fixed line ratio of $\emline{NII}/\Halpha\simeq0.41$. Predictions are shown for three different dust attenuation methods: \protect\citet{Ferrara99}, \protect\citet{Calzetti00} and \protect\citet{Charlot00}. Both the redshift distributions and the differential fluxes correspond to blended $\Halpha+\emline{NII}$ fluxes. In the right-hand column the solid grey-shaded regions correspond to fluxes below the designated flux limit.}
  \label{fig:forecasts}
\end{figure*}

\subsection{Redshift distributions}

We begin by presenting predictions for redshift distributions that we might expect for our Euclid-like and WFIRST-like survey strategies. The distributions when adopting each of the three dust methods are shown in the left-hand column of Fig.~\ref{fig:forecasts}, with the various lines showing the mean counts over our 1000 realisations and the shaded regions showing the 1$\sigma$ uncertainties. The distributions shown in Fig.~\ref{fig:forecasts} assume different flux selections, as specified in each panel, using the $\Halpha+\emline{NII}$ blended fluxes. In Table~\ref{tab:euclid_redshift_distribution} we show the mean and $1\sigma$ redshift counts for two Euclid-like survey strategies, with  $\Halpha+\emline{NII}$ blended flux limits of $2\times10^{-16}\ergPerSecondPerCM$ and  $3\times10^{-16}\ergPerSecondPerCM$. The counts for a WFIRST-like survey strategy with a flux limit of  $1\times10^{-16}\ergPerSecondPerCM$, in the instances of both a blended $\Halpha+\emline{NII}$ selection and de-blended $\Halpha$ selection, are shown in Table~\ref{tab:wfirst_redshift_distribution}. We do not show the de-blended counts for the Euclid-like surveys as the resolution of the Euclid grism means that Euclid will \emph{not} be able to de-blend the majority of the $\Halpha$ and $\emline{NII}$ lines. Note that the counts in these tables correspond to the expected numbers of galaxies that will be observed. For grism spectroscopy, the number of galaxies with redshift measurements will typically be $\sim$50 per cent of those observed, depending on the efficiency of the survey, which in turn depend on instrumentation and noise parameters.

\begin{table*}
\centering
\caption{Predicted redshift distribution of the number counts of $\Halpha$-emitting galaxies, ${\rm d}N/{\rm d}z$, per square degree for two Euclid-like $\Halpha$-selected surveys with $\Halpha+\emline{NII}$ blended flux limits of $3\times10^{-16}\ergPerSecondPerCM$ (columns 3-5) and $2\times10^{-16}\ergPerSecondPerCM$ (columns 6-8). Predicted counts are reported for each of the three dust methods: \protect\citet{Ferrara99}, \protect\citet{Calzetti00} and \protect\citet{Charlot00}. In each case we show the mean counts and $1\sigma$ uncertainty over 1000 realisations for the optical depth dust parameters.
The efficiency of each survey is instrumentation dependent, and has not been included.}
\begin{tabular}{|c|c|c|c|c|c|c|c|}
\hline
\multicolumn{2}{|c|}{Redshift}&\multicolumn{3}{|c|}{Euclid-like $\left(3\times 10^{-16}\right)$}&\multicolumn{3}{|c|}{Euclid-like $\left(2\times 10^{-16}\right)$}\\
Lower limit&Upper limit&\citeauthor{Ferrara99}&\citeauthor{Calzetti00}&\citeauthor{Charlot00}&\citeauthor{Ferrara99}&\citeauthor{Calzetti00}&\citeauthor{Charlot00}\\
&&(\citeyear{Ferrara99})&(\citeyear{Calzetti00})&(\citeyear{Charlot00})&(\citeyear{Ferrara99})&(\citeyear{Calzetti00})&(\citeyear{Charlot00})\\
\hline\hline
\input{euclidRedshiftCountsVariableNII.tex}
\hline
\end{tabular}
\label{tab:euclid_redshift_distribution}
\end{table*}

\begin{table*}
\centering
\caption{Predicted redshift distribution of the number counts of $\Halpha$-emitting galaxies, ${\rm d}N/{\rm d}z$, per square degree for a WFIRST-like $\Halpha$-selected survey with $\Halpha+\emline{NII}$ blended flux limit of $1\times10^{-16}\ergPerSecondPerCM$ (columns 3-5) and $\Halpha$ de-blended flux limit of $1\times10^{-16}\ergPerSecondPerCM$ (columns 6-8). Predicted counts are reported for each of the three dust methods: \protect\citet{Ferrara99}, \protect\citet{Calzetti00} and \protect\citet{Charlot00}. In each case we show the mean counts and $1\sigma$ uncertainty over 1000 realisations for the optical depth dust parameters.
The efficiency of each survey is instrumentation dependent, and has not been included.}
\begin{tabular}{|c|c|c|c|c|c|c|c|}
\hline
\multicolumn{2}{|c|}{Redshift}&\multicolumn{3}{|c|}{WFIRST-like ($\Halpha+\emline{NII}$ blended flux limit)}&\multicolumn{3}{|c|}{WFIRST-like ($\Halpha$ de-blended flux limit)}\\
Lower limit&Upper limit&\citeauthor{Ferrara99}&\citeauthor{Calzetti00}&\citeauthor{Charlot00}&\citeauthor{Ferrara99}&\citeauthor{Calzetti00}&\citeauthor{Charlot00}\\
&&(\citeyear{Ferrara99})&(\citeyear{Calzetti00})&(\citeyear{Charlot00})&(\citeyear{Ferrara99})&(\citeyear{Calzetti00})&(\citeyear{Charlot00})\\
\hline\hline
\input{wfirstRedshiftCountsVariableNII_combined.tex}

\hline
\end{tabular}
\label{tab:wfirst_redshift_distribution}
\end{table*}

All of the redshift distributions in Fig.~\ref{fig:forecasts} show a steady decline in the number of galaxies with increasing redshift as a result of the applied flux limits. The various peaks and troughs are most likely a result of sample variance and the finite number of galaxies caused by the finite volume of the cosmological simulation we have used. In addition, these features could be being further amplified by the periodic repetition of the simulation box used in the lightcone construction process. This demonstrates the need for larger volume simulations. We plan to apply \galacticus{} to larger volume simulations in the future. Understandably, we see that the noise features are the most pronounced in the Euclid-like strategy with flux limit of $3\times10^{-16}\ergPerSecondPerCM$, particularly towards higher redshift, where galaxy number density is lowest compared to lower redshift bins. Increasing the flux limit to $2\times10^{-16}\ergPerSecondPerCM$ and then to $1\times10^{-16}\ergPerSecondPerCM$, in the WFIRST-like strategy, leads to a smoothing of the redshift distributions as more fainter galaxies begin to be detected and the number density of detected galaxies increases. Note that in this work we have only considered a single $\Halpha$ selection and do not include galaxies that would have been detected by additional emission lines, such as \emline{OIII}, which would introduce additional peaks in the redshift distribution as lines of differing relative strength enter the wavelength ranges of the Euclid and WFIRST grisms. We will examine the impact of additional emission lines, such as \emline{OIII}, in future work.

If we focus on any one survey strategy and compare the redshift distributions from the different dust methods we can make several observations. Firstly, in all three cases the \citet{Charlot00} method shows the largest relative uncertainty in the counts, which is understandable given the degeneracy seen in Fig.~\ref{fig:dust_calibration} and Fig.~\ref{fig:monteCarloSamples}. In contrast, the \citet{Ferrara99} method shows the smallest uncertainty due to the narrow range of permissible optical depths seen in Fig.~\ref{fig:monteCarloSamples}. Secondly we see that at $z\lesssim1$ the \citet{Calzetti00} and \citet{Charlot00} methods typically show closer agreement, with the \citet{Ferrara99} method predicting between 10--20 per cent fewer galaxies. At higher redshifts, however, the \citet{Ferrara99} and \citet{Calzetti00} methods show closer agreement, though the redshift at which the \citet{Ferrara99} method begins to predict higher counts than the \citet{Charlot00} method increases with increasing flux limit, from $z\sim1.1$ for the Euclid-like $f_{\rm obs}\geqslant 3\times10^{-16}\ergPerSecondPerCM$ selection to $z\sim1.6$ for the WFIRST-like $f_{\rm obs}\geqslant 1\times10^{-16}\ergPerSecondPerCM$ selection. At high redshifts the \citet{Charlot00} method typically predicts 40--50 per cent fewer galaxies than the \citet{Ferrara99} and \citet{Calzetti00} methods.

The fact that different methods yield different counts at different redshifts reflects the changes in the relative dust attenuation strengths of the methods with redshift. In other words, if all methods showed the same strength of dust attenuation then we would expect their redshift distributions to agree perfectly and so any differences in the redshift distributions are a direct consequence of the different dust methods disagreeing on the strength of dust attenuation. Therefore, the \citet{Ferrara99} method typically predicts fewer counts at low redshifts because it typically predicts a higher dust attenuation at those redshifts compared the other two methods. In the same way, the \citet{Charlot00} method has a weaker dust attenuation at high redshift compared to the other two methods. In fact between the upper and lower redshift limit of each survey strategy, the \citet{Charlot00} method consistently shows the largest overall fractional change in number of galaxies, perhaps reflecting the greatest overall change in dust attenuation strength with redshift. In contrast the \citet{Ferrara99} method typically shows the smallest fractional change in the number of galaxies, which might suggest that this method shows the weakest overall change in dust attenuation strength with redshift. The variation in the redshifts at which pairs of dust methods agree on the galaxy counts suggests an additional dependence on luminosity, i.e. the relative attenuation strength predicted by each of the different methods changes with redshift and luminosity. This is supported by the discussion in Section~\ref{sec:luminosity_functions}, where we saw disagreements between the $\Halpha$ luminosity functions predicted by the different dust methods.

\subsection{Flux counts}

The right-hand column of Fig.~\ref{fig:forecasts} shows the cumulative flux counts for the three different dust methods. We see that the \citet{Ferrara99} method consistently has a shallower slope as a function of flux, leading to this method displaying an excess of bright galaxies compared to the \citet{Calzetti00} and \citet{Charlot00} methods. The slopes for the \citet{Calzetti00} and \citet{Charlot00} methods are typically in much closer agreement, with the \citet{Charlot00} method predicting fewer galaxies for fluxes fainter than approximately $1\times10^{-15}\ergPerSecondPerCM$. By flux limits of $1\times10^{-16}\ergPerSecondPerCM$ there is a closer agreement between the three dust methods, with the \citet{Calzetti00} and \citet{Charlot00} methods now predicting a slightly higher number of galaxies compared to the \citet{Ferrara99} method. The counts for all three of the dust methods tentatively show a break, with the slope becoming shallower for fluxes brighter than approximately $3\times10^{-15}\ergPerSecondPerCM$. As with our examination of the redshift counts, the similarities and differences in the flux counts from the three different methods again most likely reflects the relative strength in dust attenuation between the three methods as a function of luminosity. For reference, tabulated differential flux counts for the Euclid-like and WFIRST-like strategies are provided in Appendix~\ref{sec:differentialCounts}.

The cumulative blended flux counts as predicted by the three empirical models from \citet{Pozzetti16} are also shown for comparison in the right-hand column of Figure~\ref{fig:forecasts}. Note that for the \citet{Pozzetti16} models we do not have stellar mass or star formation rate information and so cannot model \emline{NII} contamination using our SDSS catalogue (see Section~\ref{sec:modellingNII}). Instead we simply introduce \emline{NII} contamination into the \citet{Pozzetti16} counts using a fixed $\emline{NII}/\Halpha$ ratio of $\emline{NII}/\Halpha\simeq0.41$.  We can clearly see that for fluxes fainter than approximately $4\times10^{-16}\ergPerSecondPerCM$ the scatter between the \galacticus{} counts from the three dust methods is smaller than scatter between the counts from the three \citet{Pozzetti16} models. At the flux limits for each of the Euclid-like and the WFIRST-like observing strategies \citet{Pozzetti16} models one and two predict an excess of galaxies, by approximately a factor of 2, compared to \citet{Pozzetti16} model 3. The \galacticus{} counts from the three different dust methods are in much closer agreement with each other, and are typically most consistent with \citet{Pozzetti16} model 3. This is not surprising, as the \citet{Pozzetti16} model 3 is also calibrated using WISP data. Towards brighter fluxes the counts from the \citet{Pozzetti16} models fall off much more rapidly than the \galacticus{} counts, such that by approximately $1\times10^{-15}\ergPerSecondPerCM$ all three of the \citet{Pozzetti16} models predict counts that are lower than \galacticus{}. At $1\times10^{-15}\ergPerSecondPerCM$ the scatter between counts from the different \galacticus{} dust models has grown to a factor of approximately 2, compared to a scatter of approximately a factor of 5 between the three \citet{Pozzetti16} models. We note that the scatter in the counts from the \citet{Pozzetti16} models is sensitive to the particular observational datasets and parametrization choices used in their fitting. We could tentatively argue that our \galacticus{} counts are more predictive since \galacticus{} is a physical model of galaxy formation with all of the galaxy properties calculated self-consistently. However, we acknowledge that \galacticus{} still requires a variety of observational datasets for proper calibration and so the scatter in the \galacticus{} predictions will still, to some extent, be moderated by the scatter in the observations used in that calibration. As such, we stress that \galacticus{} would need to be calibrated using the same observational datasets as were used by \citeauthor{Pozzetti16} before a more rigorous comparison can be made.

\begin{figure}
  \centering
  \includegraphics[width=0.48\textwidth]{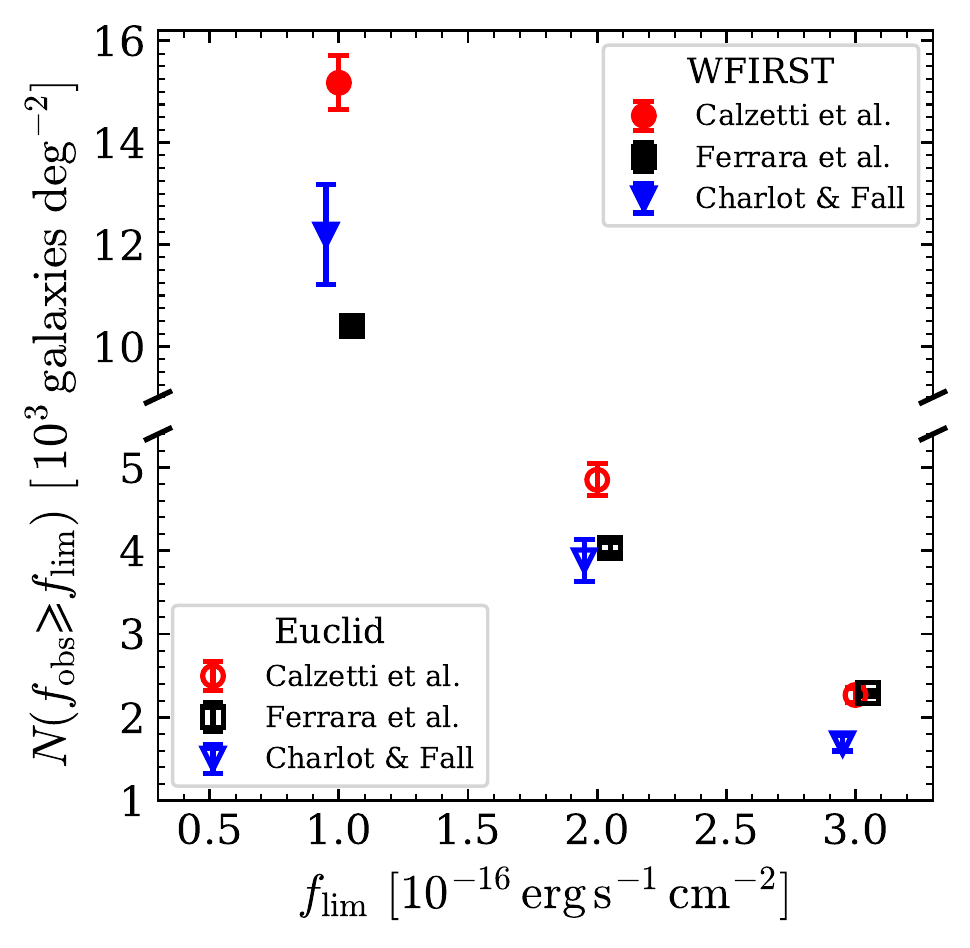}
  \caption{\galacticus{} predictions for the cumulative number of $\Halpha$-emitting galaxies visible above a specified flux limit for a Euclid-like survey (empty points) and WFIRST-like survey (filled points). The points for the \protect\citet{Ferrara99} and  \protect\citet{Charlot00} methods have been offset in the x-direction for clarity. We assume that the fluxes are blended $\Halpha+\emline{NII}$ fluxes. For each of the three dust methods, the points correspond to the mean counts from 1000 Monte-Carlo realisations (see text for details). The error bars correspond to the $\pm1\sigma$ uncertainty. The mean values and uncertainties are listed in Table~\ref{tab:cumulativeFluxCounts}.} 
  \label{fig:cumulativeForecasts}
\end{figure}

\begin{table*}
\centering
\caption{Predicted cumulative number of $\Halpha$-emitting galaxies per square degree for a Euclid-like $\Halpha$-selected galaxy survey (columns (2-4)
and a WFIRST-like $\Halpha$-selected galaxy survey (columns 5-7)
In each case, the redshift ranges over which galaxies were selected are shown in the table. The upper half of the table shows the counts for $\Halpha+\emline{NII}$ blended fluxes, whilst the lower half of the table shows the counts for  $\Halpha$ de-blended fluxes. Note that for the Euclid-like survey we only show counts for blended fluxes. Predicted counts are reported for three dust methods: \protect\citet{Ferrara99}, \protect\citet{Calzetti00} and \protect\citet{Charlot00}. Note that the flux limits correspond to $\Halpha+\emline{NII}$ blended fluxes. The efficiency of each survey is instrumentation dependent, and has not been included. On the last row of the table we show again the $\chi^2_{\rm min}/\nu$ values from Table~\ref{tab:chiSquaredParameters}.}
\begin{tabular}{|c|c|c|c|c|c|c|}
\hline
Flux limit&\multicolumn{3}{|c|}{Euclid-like ($0.9\leqslant z\leqslant1.8$)}&\multicolumn{3}{|c|}{WFIRST-like ($1\leqslant z\leqslant 2$)}\\
$\ergPerSecondPerCM$&\citeauthor{Ferrara99}&\citeauthor{Calzetti00}&\citeauthor{Charlot00}&\citeauthor{Ferrara99}&\citeauthor{Calzetti00}&\citeauthor{Charlot00}\\
&(\citeyear{Ferrara99})&(\citeyear{Calzetti00})&(\citeyear{Charlot00})&(\citeyear{Ferrara99})&(\citeyear{Calzetti00})&(\citeyear{Charlot00})\\
\hline\hline
\multicolumn{7}{|c|}{\textsc{Cumulative counts from $\Halpha+\emline{NII}$ blended fluxes}}\\
\input{forecastsCumulativeFluxCountsVariableNII.tex}
\hline
\multicolumn{7}{|c|}{\textsc{Cumulative counts from $\Halpha$ de-blended fluxes}}\\
\input{forecastsCumulativeFluxCountsVariableNoNII.tex}
\hline\hline
$\chi^2_{\rm min}/\nu$&7.9&2.9&6.0&7.9&2.9&6.0\\
\hline\hline
\end{tabular}
\label{tab:cumulativeFluxCounts}
\end{table*}

The cumulative number counts at the blended flux limits for the Euclid-like and WFIRST-like strategies are summarised in Fig.~\ref{fig:cumulativeForecasts}, as well as in Table~\ref{tab:cumulativeFluxCounts} where for the WFIRST-like strategy we also show the cumulative counts for the $\Halpha$ de-blended fluxes. We remind the reader that de-blended counts are not shown for the Euclid-like strategy as it is expected that the Euclid grism will be unable to resolve the majority of the $\Halpha$ and \emline{NII} lines.

For the Euclid-like strategies, over the redshift range $0.8\leqslant z\leqslant 1.9$, we find approximately 3900--4800 galaxies per square degree for a flux limit of $2\times10^{-16}\ergPerSecondPerCM$ and approximately 1700--2300 galaxies per square degree for a brighter flux limit of $3\times10^{-16}\ergPerSecondPerCM$. Use of the different dust methods leads to scatter of between 2--30 per cent in the counts. We see that for both of the Euclid-like strategies in Fig.~\ref{fig:cumulativeForecasts}, two of the three dust methods predict cumulative counts that are in good agreement. At the blended flux limit of $2\times10^{-16}\ergPerSecondPerCM$ we see good agreement between the counts from the \citet{Ferrara99} and \citet{Charlot00} methods, which are consistent within uncertainty with each other, whilst the \citet{Calzetti00} method predicts approximately 20 per cent more galaxies per square degree. Considering the brighter limit of $3\times10^{-16}\ergPerSecondPerCM$, the \citet{Ferrara99} and \citet{Calzetti00} methods now show excellent agreement, whilst the \citet{Charlot00} method predicts approximately 26 per cent fewer galaxies per square degree. Compared to the Euclid-like strategies, counts predicted for the the WFIRST-like strategy show a larger spread, with none of the dust methods agreeing within $1\sigma$ uncertainty. For a blended flux limit of $1\times10^{-16}\ergPerSecondPerCM$ the counts range from about 15200 galaxies per square degree for the \citet{Calzetti00}, to about 10400 galaxies per square degree for the \citet{Ferrara99} method.  The scatter in the counts for the WFIRST-case is between 20--50 per cent, though for a fainter blended flux limit of $2\times10^{-16}\ergPerSecondPerCM$ the scatter in the WFIRST-like counts decreases slightly to between 20--40 per cent, with count predictions comparable to those for the Euclid-like case with the same flux limit. For the WFIRST-like strategy, if we compare the blended and de-blended flux counts shown in Table~\ref{tab:cumulativeFluxCounts} we see that for the WFIRST flux limit of $1\times10^{-16}\ergPerSecondPerCM$ de-blending of the $\Halpha$ and \emline{NII} lines leads to approximately 30 per cent fewer galaxies per square degree. Following de-blending, the number of galaxies brighter than $1\times10^{-16}\ergPerSecondPerCM$ is between approximately 7500--10600 galaxies per square degree, with a scatter between 30--40 per cent.


\section{Summary \& Conclusions}
\label{sec:conclusions}

Assessing the performance of the Euclid and WFIRST missions requires estimates of the expected number density of $\Halpha$-emitters as a function of redshift. In this work we have used an open source semi-analytical galaxy formation model, \galacticus{}, to predict the number counts and redshift distributions of $\Halpha$-emitting galaxies for a Euclid-like and a WFIRST-like survey strategy.

We construct a 4 square degree lightcone catalogue by processing the dark matter merger trees of the Millennium Simulation with the \galacticus{} model. Emission lines are modelled in \galacticus{} by interpolating over a library of emission line luminosities obtained from the \cloudy{} code. The emission line luminosities are then processed to incorporate attenuation due to interstellar dust, which can be modelled using several different methods. Here we consider three dust methods from \citet{Ferrara99}, \citet{Charlot00} and \citet{Calzetti00}, though we stress that any user-specified dust method can be used in conjunction with \galacticus{}. 

Before making predictions for Euclid-like and WFIRST-like surveys, we first compare the \galacticus{} galaxy counts to existing observed $\Halpha$-emitting galaxy counts over the redshift range $0.7\leqslant z\leqslant 1.5$ from the WISP survey, as presented by \citet{Mehta15}. Since we are comparing to counts from $\Halpha+\emline{NII}$ blended fluxes, we introduce \emline{NII} into the \galacticus{} $\Halpha$ fluxes by using the SDSS catalogue from \citet{Masters16} to assign the \galacticus{} galaxies $\emline{NII}/\Halpha$ line ratios based upon a nearest-neighbour search in stellar mass versus specific star formation rate space.

By minimising the $\chi^2$ statistic we are able to identify optical depth parameters for each dust method that yield \galacticus{} counts that are broadly consistent with the WISP counts, particularly at flux limits fainter than $2\times 10^{-16}\ergPerSecondPerCM$, which are comparable to the flux depths of our adopted Euclid-like and WFIRST-like survey strategies. However, the large reduced $\chi^2$ values, $\chi^2/\nu\gtrsim3$, suggest that further calibration of \galacticus{} or revision of the dust methods is required for future analyses. The \citet{Calzetti00} method provides the best overall fit to the WISP data, despite being the simplest and least physically motivated dust attenuation method. The poorest fit is from the \citet{Ferrara99} method, which is the most physically motivated of the three methods, with the ISM optical depths being computed directly from various galaxy properties that are output directly from \galacticus{}. This direct computation adds further constraint to the \citeauthor{Ferrara99} method, reducing the flexibility of this method compared to the more parametrised methods of \citet{Charlot00} and \citet{Calzetti00}. The optical depths that we identify suggest that only weak dust attenuation, particularly in molecular clouds, is required to match the WISP counts. Comparing the optical depths of the \galacticus{} galaxies with optical depths measured from WISP we find that the model and observations are consistent within uncertainty (though the \galacticus{} optical depths are slightly smaller). 

We have also briefly examined how adopting a fixed \emline{NII} contamination of 29 per cent, consistent with the value adopted in the WISP analyses, impacts the counts and required dust strength. We have also estimated the impact of cosmic variance, which we expect to have little effect on the predicted number counts at flux limits fainter than $2\times 10^{-16}\ergPerSecondPerCM$, comparable with the flux limits for Euclid and WFIRST. We note, however, that the lightcone analyses that we have carried out are time-consuming and expensive in computing resources. As such we have limited our lightcone size to 4 square degrees for this study in order to provide timely input to Euclid and WFIRST projects. We plan to build significantly larger lightcones in future work.

To further check that our calibrated optical depth parameters are reasonable we also compare the \galacticus{} predictions for the $\Halpha$ luminosity function to observational estimates from WISP \citep{Colbert13} and from HiZELS \citep{Sobral13}. The \galacticus{} luminosity function has a shape and normalisation consistent with the luminosity functions from WISP and the lowest redshift bin from HiZELS. However, at higher redshift the \galacticus{} luminosity function becomes a progressively worse fit to the HiZELS observations. This disagreement could suggest that the dust methods may be lacking some additional redshift evolution or dependence on other galaxy properties, or that \galacticus{} requires further calibration. Investigating these possibilities requires rigorous calibration of the \galacticus{} model, which we leave for future work. 

Finally, we use the \galacticus{} lightcone to present predictions for the redshift distribution and cumulative flux counts for two Euclid-like surveys and a WFIRST-like survey. To marginalise over the choice of optical depth dust parameters we use $\chi^2$ as a function of optical depth to construct likelihoods with which we randomly sample 1000 realisations for the parameters for each dust method. For each realisation we re-compute the Euclid-like and WFIRST-like flux and redshift counts. The predictions that we present correspond to the mean and $1\sigma$ uncertainty from these 1000 realisations. We compare our cumulative counts to the counts from \citet{Pozzetti16}, which we correct to introduce \emline{NII} contamination using $\emline{NII}/\Halpha\simeq0.41$, and find that the \galacticus{} counts are most consistent with \citet{Pozzetti16} model three. For a Euclid-like survey with redshift range $0.9\leqslant z\leqslant 1.8$ and $\Halpha+\emline{NII}$ blended flux limit of $3\times 10^{-16}\ergPerSecondPerCM$ we predict a galaxy number density between 1700--2300 galaxies per square degree. Considering a fainter blended flux limit of $2\times 10^{-16}\ergPerSecondPerCM$ increases the number densities to between 3900--4800 galaxies per square degree. The scatter in these counts is between 2--30 per cent. For a WFIRST-like survey with redshift range $1\leqslant z\leqslant 2$ and blended flux limit of $1\times 10^{-16}\ergPerSecondPerCM$ we predict a number density between 10400-15200 galaxies per square degree. The WFIRST-like counts have a slightly larger scatter of 20--50 per cent. For the WFIRST-like survey, we find $\Halpha$ de-blended counts of 7500--10600 galaxies per square degree. 

Note that all the $\Halpha$-emitter counts discussed in this paper are expected number counts of target galaxies for spectroscopy, and \emph{not} the counts of galaxies with redshift measurements. The latter will depend on the redshift purity and completeness for each survey, which in turn depends on instrumentation and noise parameters.

In future work we plan to further exploit the \galacticus{} model to examine a variety of other properties of emission line galaxies, including the distribution of \emline{OIII} luminosities fluxes and the contamination from \emline{NII}.

\section*{Acknowledgements}
We thank James Colbert, Ivano Baronchelli, Shoubaneh Hemmati and Kirsten Larson for many helpful conversations. We also thank the anonymous referee for their valuable and constructive comments that helped greatly improve the manuscript. AM acknowledges sponsorship of a NASA Postdoctoral Program Fellowship. AM, AK and JR were supported by JPL, which is run under contract by California Institute of Technology for NASA. This work was supported by NASA ROSES grant 12-EUCLID12-0004 and by NASA grant 15-WFIRST15-0008 “Cosmology with the High Latitude Survey” WFIRST Science Investigation Team (SIT). Copyright 2017. All rights reserved.


\bibliographystyle{mn2e_trunc8}
\bibliography{aim}

\appendix

\section{Differential flux counts}
\label{sec:differentialCounts}

\begin{table*}
\centering
\caption{Predicted differential flux counts, ${\rm d}N/{\rm d}\log_{10}f$, per square degree for a Euclid-like $\Halpha$-selected galaxy survey over the redshift range $0.9\leqslant z \leqslant 1.8$. The counts correspond to $\Halpha+\emline{NII}$ blended fluxes. Predicted counts are reported for three dust methods: \protect\citet{Ferrara99}, \protect\citet{Calzetti00} and \protect\citet{Charlot00}. The efficiency of each survey is instrumentation dependent, and has not be included.}
\begin{tabular}{|c|c|c|c|c|}
\hline
\multicolumn{2}{|c|}{flux [$10^{-16}\ergPerSecondPerCM$]}&\multicolumn{3}{|c|}{Euclid-like ($0.9\leqslant z \leqslant 1.8$)}\\
Lower limit&Upper limit&\citet{Ferrara99}&\citet{Calzetti00}&\citet{Charlot00}\\
\hline\hline
\input{euclidDifferentialCountsVariableNII.tex}
\hline
\end{tabular}
\label{tab:euclidDifferentialFluxCounts}
\end{table*}

\begin{table*}
\centering
\caption{Predicted differential flux counts, ${\rm d}N/{\rm d}\log_{10}f$, per square degree for a WFIRST-like $\Halpha$-selected galaxy survey over the redshift range $1\leqslant z \leqslant 2$. The counts presented in columns 3-5 correspond to $\Halpha+\emline{NII}$ blended fluxes, whilst the counts presented in columns 6-8 correspond to $\Halpha$ de-blended fluxes. Predicted counts are reported for three dust methods: \protect\citet{Ferrara99}, \protect\citet{Calzetti00} and \protect\citet{Charlot00}. The efficiency of each survey is instrumentation dependent, and has not be included.}
\begin{tabular}{|c|c|c|c|c|c|c|c|c|c|c|}
\hline
\multicolumn{2}{|c|}{flux [$10^{-16}\ergPerSecondPerCM$]}&\multicolumn{3}{|c|}{WFIRST-like ($\Halpha+\emline{NII}$ blended)}&\multicolumn{3}{|c|}{WFIRST-like ($\Halpha$ de-blended)}\\
Lower limit&Upper limit&\citeauthor{Ferrara99}&\citeauthor{Calzetti00}&\citeauthor{Charlot00}&\citeauthor{Ferrara99}&\citeauthor{Calzetti00}&\citeauthor{Charlot00}\\
&&(\citeyear{Ferrara99})&(\citeyear{Calzetti00})&(\citeyear{Charlot00})&(\citeyear{Ferrara99})&(\citeyear{Calzetti00})&(\citeyear{Charlot00})\\
\hline\hline
\input{wfirstDifferentialCountsVariableNII.tex}
\hline
\end{tabular}
\label{tab:wfirstDifferentialFluxCounts}
\end{table*}

In Table~\ref{tab:euclidDifferentialFluxCounts} we show the predicted tabulated differential blended flux counts for the Euclid-like strategies, with blended flux limits of $3\times10^{-16}\ergPerSecondPerCM$ and $2\times10^{-16}\ergPerSecondPerCM$ over the redshift range $0.9\leqslant z \leqslant 1.8$. Counts are shown for each of the three dust methods applied to the \galacticus{} lightcone: \citet{Ferrara99, Calzetti00} and \citet{Charlot00}. Equivalent differential flux counts for the WFIRST-like strategy, assuming a blended flux limit of $1\times10^{-16}\ergPerSecondPerCM$ over the redshift range $1\leqslant z \leqslant 2$, are shown in Table~\ref{tab:wfirstDifferentialFluxCounts}. For the WFIRST-like strategy we show the counts for both the blended $\Halpha+\emline{NII}$ fluxes and the $\Halpha$ de-blended fluxes. De-blended counts are not shown for the Euclid-like strategy as it is expected that the Euclid grism will be unable to resolve the majority of the $\Halpha$ and \emline{NII} lines.

\end{document}

%% file: chiSquaredDustParameters.tex
Ferrara et al. (1999)&7.9&$\tau^{\rm MC}_{\rm V}=0.00^{+0.006}_{-0.00}$\\
Calzetti et al. (2000)&2.9&$\tau_{\rm V}=0.64^{+0.03}_{-0.02}$\\
Charlot \& Fall (2000)&6.0&$\tau^{\rm ISM}_{\rm V}=0.032^{+0.002}_{-0.008}$\\
&&$\tau^{\rm MC}_{\rm V}=0.0^{+0.1}_{-0.0}$\\

%% file: euclidRedshiftCountsVariableNII.tex
0.9&1&$3598\pm50$&$4493\pm195$&$4240\pm243$&$7716\pm103$&$10057\pm361$&$10535\pm472$\\
1&1.1&$3050\pm40$&$3546\pm169$&$3083\pm200$&$6229\pm100$&$8500\pm344$&$8014\pm523$\\
1.1&1.2&$2467\pm44$&$2582\pm114$&$1964\pm121$&$4475\pm72$&$6323\pm273$&$4998\pm472$\\
1.2&1.3&$2537\pm45$&$2409\pm86$&$1689\pm81$&$4225\pm64$&$5463\pm260$&$3931\pm374$\\
1.3&1.4&$2690\pm43$&$2371\pm76$&$1623\pm78$&$4296\pm71$&$5031\pm221$&$3455\pm246$\\
1.4&1.5&$2193\pm27$&$1926\pm55$&$1227\pm60$&$3384\pm51$&$3690\pm148$&$2446\pm178$\\
1.5&1.6&$2238\pm33$&$1890\pm60$&$1093\pm58$&$3500\pm60$&$3387\pm106$&$2078\pm141$\\
1.6&1.7&$2158\pm31$&$1775\pm49$&$1006\pm49$&$3365\pm51$&$3085\pm106$&$1747\pm113$\\
1.7&1.8&$1951\pm23$&$1627\pm43$&$943\pm44$&$3044\pm44$&$2793\pm77$&$1524\pm108$\\

%% file: wfirstRedshiftCountsVariableNII_combined.tex
1.0&1.1&$20450\pm210$&$25514\pm672$&$27901\pm850$&$14507\pm155$&$19120\pm545$&$20486\pm725$\\
1.1&1.2&$15094\pm173$&$21723\pm652$&$21359\pm1049$&$10410\pm123$&$15703\pm565$&$14923\pm1004$\\
1.2&1.3&$13400\pm165$&$21049\pm668$&$18453\pm1485$&$9270\pm143$&$14771\pm592$&$12527\pm1352$\\
1.3&1.4&$12329\pm161$&$20335\pm741$&$16041\pm1842$&$8664\pm137$&$13864\pm587$&$10595\pm1458$\\
1.4&1.5&$9039\pm137$&$15077\pm581$&$10596\pm1533$&$6348\pm98$&$10152\pm442$&$6906\pm1095$\\
1.5&1.6&$8253\pm132$&$13377\pm531$&$8421\pm1270$&$5987\pm90$&$8827\pm395$&$5487\pm802$\\
1.6&1.7&$7517\pm111$&$11191\pm486$&$6653\pm958$&$5607\pm83$&$7307\pm323$&$4249\pm558$\\
1.7&1.8&$6617\pm101$&$9174\pm407$&$5169\pm670$&$5025\pm74$&$6035\pm252$&$3409\pm419$\\
1.8&1.9&$6101\pm105$&$7865\pm323$&$4166\pm520$&$4697\pm68$&$5305\pm201$&$2834\pm311$\\
1.9&2.0&$5361\pm78$&$6462\pm228$&$3309\pm386$&$4215\pm72$&$4522\pm135$&$2311\pm216$\\

%% file: forecastsCumulativeFluxCountsVariableNII.tex
$1\times 10^{-16}$&N/A&N/A&N/A&$10403\pm138$&$15176\pm528$&$12195\pm987$\\
$2\times 10^{-16}$&$4036\pm62$&$4849\pm192$&$3884\pm252$&$3797\pm60$&$4307\pm170$&$3059\pm230$\\
$3\times 10^{-16}$&$2291\pm34$&$2267\pm85$&$1688\pm91$&$2263\pm33$&$2084\pm73$&$1414\pm75$\\

%% file: forecastsCumulativeFluxCountsVariableNoNII.tex
$1\times 10^{-16}$&N/A&N/A&N/A&$7467\pm102$&$10566\pm402$&$8365\pm765$\\
$2\times 10^{-16}$&N/A&N/A&N/A&$2827\pm42$&$2877\pm108$&$2025\pm140$\\
$3\times 10^{-16}$&N/A&N/A&N/A&$1771\pm24$&$1479\pm46$&$1032\pm45$\\

%% file: euclidDifferentialCountsVariableNII.tex
1&2&N/A&N/A&N/A\\
2&3&$9908\pm175$&$14746\pm579$&$12527\pm941$\\
3&4&$5527\pm72$&$7241\pm330$&$5516\pm396$\\
4&5&$3980\pm53$&$4328\pm170$&$3025\pm189$\\
5&6&$3043\pm40$&$2873\pm124$&$2071\pm96$\\
6&7&$2590\pm43$&$2238\pm60$&$1588\pm79$\\
7&8&$2253\pm44$&$1918\pm27$&$1346\pm53$\\
8&9&$1907\pm32$&$1659\pm41$&$1233\pm38$\\
9&10&$1720\pm23$&$1461\pm46$&$1110\pm38$\\
10&20&$1212\pm16$&$846\pm32$&$630\pm31$\\
20&30&$509\pm3$&$257\pm9$&$254\pm16$\\
30&40&$225\pm2$&$119\pm8$&$121\pm12$\\
40&50&$147\pm1$&$34\pm5$&$84\pm10$\\
50&60&$60\pm0$&$17\pm2$&$27\pm8$\\
60&70&$31\pm0$&$9\pm1$&$4\pm4$\\
70&80&$23\pm1$&$2\pm0$&$3\pm2$\\
80&90&$7\pm0$&$2\pm0$&$1\pm1$\\
90&100&$2\pm0$&$2\pm0$&$0\pm1$\\

%% file: wfirstDifferentialCountsVariableNII.tex
1&2&$22046\pm260$&$36193\pm1198$&$30484\pm2521$&$15486\pm201$&$25636\pm977$&$21184\pm2092$\\
2&3&$8720\pm165$&$12696\pm523$&$9412\pm904$&$6044\pm109$&$7972\pm369$&$5699\pm551$\\
3&4&$5260\pm76$&$6373\pm278$&$4174\pm286$&$4031\pm62$&$4067\pm126$&$2635\pm146$\\
4&5&$3982\pm63$&$3899\pm134$&$2517\pm135$&$3097\pm60$&$2777\pm79$&$1837\pm78$\\
5&6&$3198\pm36$&$2777\pm100$&$1827\pm80$&$2529\pm36$&$2097\pm37$&$1415\pm48$\\
6&7&$2715\pm38$&$2277\pm51$&$1464\pm57$&$2108\pm23$&$1801\pm51$&$1257\pm36$\\
7&8&$2322\pm41$&$1939\pm26$&$1293\pm39$&$1838\pm19$&$1599\pm65$&$1168\pm36$\\
8&9&$1973\pm27$&$1645\pm40$&$1205\pm38$&$1687\pm15$&$1358\pm44$&$1033\pm50$\\
9&10&$1779\pm20$&$1440\pm48$&$1091\pm47$&$1540\pm17$&$1166\pm35$&$868\pm58$\\
10&20&$1185\pm15$&$797\pm33$&$588\pm44$&$931\pm9$&$595\pm22$&$461\pm41$\\
20&30&$458\pm2$&$213\pm10$&$201\pm19$&$321\pm4$&$141\pm7$&$140\pm18$\\
30&40&$171\pm2$&$84\pm5$&$79\pm11$&$137\pm1$&$51\pm4$&$62\pm8$\\
40&50&$110\pm0$&$24\pm4$&$51\pm8$&$64\pm0$&$14\pm2$&$18\pm7$\\
50&60&$40\pm0$&$12\pm1$&$14\pm7$&$27\pm0$&$4\pm0$&$5\pm2$\\
60&70&$22\pm0$&$8\pm1$&$2\pm4$&$11\pm0$&$1\pm0$&$2\pm0$\\
70&80&$20\pm1$&$2\pm0$&$3\pm0$&$3\pm0$&$0\pm0$&$0\pm0$\\
80&90&$7\pm0$&$0\pm0$&$1\pm0$&$2\pm0$&$0\pm0$&$0\pm0$\\
90&100&$0\pm0$&$0\pm0$&$0\pm0$&$2\pm0$&$0\pm0$&$0\pm0$\\